\pdfoutput=1
\documentclass[12pt,a4paper]{article}
\usepackage{xcolor}
\usepackage{enumerate}
\usepackage{framed}
\usepackage{mdframed}
\usepackage{amsmath}
\usepackage{amsfonts}
\usepackage{mathrsfs}
\usepackage{amssymb}
\usepackage{graphicx, rotating}
\usepackage{epsfig}
\usepackage{latexsym}
\usepackage{graphicx}
\usepackage{color}
\usepackage{amsmath,bm,amssymb}
\usepackage{cite}
\usepackage{slashed}
\usepackage{hyperref}
\usepackage{epstopdf}
\usepackage[font=footnotesize,labelfont=]{caption}
\AppendGraphicsExtensions{.tif}
\hypersetup{colorlinks=true, citecolor=bluscuro, linkcolor=black, urlcolor=bluscuro}
\definecolor{rossos}{cmyk}{0,1,1,0.55}
\definecolor{bluscuro}{rgb}{0.15, 0.2, .85}
\definecolor{bluchiaro}{cmyk}{1,.3,0.,0.1}
\newcommand{\x}{\vec{x}}
\def\0{\vec{0}}
\newcommand{\df}{h}

\def\calo{{\cal O}}

\def\t{\eta}

\def\del{\partial}

\def\d{{\rm d}}
\def\vx{{\vec{x}}}
\def\vW{{\vec{W}}}
\def\vk{{\vec{k}}}

\def\vq{{\vec{q}}}

\def\beq{\begin{equation}}
\def\eeq{\end{equation}}

\newcommand{\D}{\Delta}

\newcommand{\expect}[1]{\left\langle #1 \right\rangle}

\begin{document}
\def\thefootnote{\fnsymbol{footnote}}

\begin{center}
\Large{\textbf{Inflation and Conformal Invariance:\\ The  Perspective from   Radial Quantization  }} \\[0.5cm]
\end{center}
\vspace{0.5cm}

\begin{center}

\large{Alex Kehagias$^{\rm a,b}$  and  Antonio Riotto$^{\rm c}$}
\\[0.5cm]

\small{
\textit{$^{\rm a}$Physics Division, National Technical University of Athens, 15780 Zografou Campus, Athens, Greece}}

\vspace{.2cm}

\small{
\textit{$^{\rm b}$Theoretical Physics Department, CERN, CH-1211 Geneva 23, Switzerland}}

\vspace{.2cm}

\small{
\textit{$^{\rm c}$Department of Theoretical Physics and Center for Astroparticle Physics (CAP) \\
24 quai E. Ansermet, CH-1211 Geneva 4, Switzerland}}

\vspace{.2cm}

\end{center}

\vspace{.7cm}

\hrule \vspace{0.3cm}
\noindent \small{\textbf{Abstract}\\ 
According to the dS/CFT correspondence,  correlators of fields generated during a primordial de Sitter phase are  constrained by three-dimensional conformal invariance. 
Using the properties of  radially quantized conformal field theories and the operator-state correspondence, we  glean  information on  some points. The Higuchi bound on the masses of spin-$s$ states in de Sitter 
 is a direct consequence of  reflection positivity in radially quantized CFT$_3$  and the fact that  scaling dimensions of operators are energies of states. The partial massless states  appearing in de Sitter   correspond from  the boundary CFT$_3$ perspective to boundary states with     highest weight for the
conformal group. We discuss  inflationary consistency relations and the role of asymptotic symmetries which 
  transform asymptotic vacua to new physically inequivalent vacua by generating long perturbation modes.  We  show that  on the 
CFT$_3$ side, asymptotic symmetries have a nice quantum mechanics interpretation. For instance, acting with  the asymptotic dilation symmetry corresponds to   evolving  states  forward (or backward) in ``time"  and  
the charge generating the asymptotic  symmetry transformation is    the Hamiltonian itself. Finally, we investigate the symmetries of  anisotropic  inflation and show that   correlators of four-dimensional free scalar fields can be reproduced  in the dual picture by considering  an isotropic  three-dimensional boundary enjoying dilation symmetry, but with a nonvanishing  vacuum expectation value  of the boundary stress-energy momentum tensor.  }

\vspace{0.3cm}
\noindent
\hrule
\def\thefootnote{\arabic{footnote}}
\setcounter{footnote}{0}

%
%
%
%


 \def\vx{\vec{ x}} 
\def\vk{\vec{k}}
\def\vy{\vec{y}}

\numberwithin{equation}{section}

\def\la{~\mbox{\raisebox{-.6ex}{$\stackrel{<}{\sim}$}}~}
\def\ga{~\mbox{\raisebox{-.6ex}{$\stackrel{>}{\sim}$}}~}
\def\bq{\begin{quote}}
\def\eq{\end{quote}}
\def\PL{{ \it Phys. Lett.} }
\def\PRL{{\it Phys. Rev. Lett.} }
\def\NP{{\it Nucl. Phys.} }
\def\PR{{\it Phys. Rev.} }
\def\MPL{{\it Mod. Phys. Lett.} }
\def\IJMP{{\it Int. J. Mod .Phys.} }
\font\tinynk=cmr6 at 10truept
\newcommand{\be}{\begin{eqnarray}}
\newcommand{\ee}{\end{eqnarray}}
\newcommand{\n}{{\bf n}}
\newcommand{\arXiv}[2]{\href{http://arxiv.org/pdf/#1}{{\tt [#2/#1]}}}
\newcommand{\arXivold}[1]{\href{http://arxiv.org/pdf/#1}{{\tt [#1]}}}

\section{Introduction \label{sec:intro}} 
Over the last decade we have  become more and more convinced that  
the cosmic microwave background anisotropies and the    large-scale structure of the universe   have been originated from some seeds
generated in the very early universe.  The leading  paradigm to  explain these  primordial fluctuations 
is  inflation \cite{inf}. 
During a stage of accelerated expansion, dubbed de Sitter (dS) period,  quantum fluctuations are stretched to cosmological scales  and, upon  horizon re-entry and thanks  to the phenomenon of  gravitational instability,   they  provide the seeds for the structures of the  universe.

Understanding  inflation, its dynamics and observational predictions is therefore of extreme importance.  From this point of view, symmetries may be of much help. It is well known that they play a crucial role in high energy physics and they  have proved to be extremely useful in cosmology too.
For instance, the consistency relations for single-field models of inflation \cite{cr,cz}, which  allow to express the squeezed limit of the $(n+1)$-point correlators in terms of $n$-point functions  and whose violation would  rule out single-field models of inflation, 
 are derived from   non-linearly realized symmetries 
of inflation corresponding to conformal transformations  of  spatial ${\mathbb R}^3$ slices. Similarly, late-time universe consistency relations for the fluctuations in the dark matter density or in the number density of galaxies  have also been recently obtained \cite{noi1,PietroniPeloso1,noi2,cr1,cr2} based on symmetry arguments. 

A crucial step to deepen our knowledge about the properties of the inflationary fluctuations has been taken by Strominger who has formulated
the so-called dS/CFT correspondence (CFT standing for conformal field theory) \cite{strominger}. It is based on the fact that 
the  de Sitter isometry SO(1,4) group acts  as conformal group 
on $\mathbb{R}^3$ when the fluctuations are on super-Hubble scales. Bulk fields $\phi(\vx,\eta)$ evolving in four-dimensional de Sitter spacetime 
and behaving  near the boundary $\eta=0$ ($\eta$ being the conformal time) as $\phi(\vx,\eta)\sim (-\eta)^\Delta\phi(\vx)$ correspond   on the three-dimensional space, where the symmetry is CFT$_3$,    to a field of conformal weight  $h=\Delta-3$.
Correlators of fields excited during a de Sitter phase are therefore  expected to be constrained by conformal invariance and the literature on this topic   is rich  and diverse \cite{anton,c0,c00,c1,c2,c3,c4,c5,c6,c7,c8,c9,c10,c11,c12,c13,c14,c15,c16,c17,c18,c19,bau,c20,sloth,c21}. In particular,
much attention has been devoted to the relation among the consistency relations,  soft theorems and the  asymptotic symmetries of de Sitter  which transform asymptotic vacua to new physically inequivalent vacua by generating  long perturbation modes \cite{c21,sloth}.

The isometries of de Sitter   impose also stringent lower bounds on the masses   of particles with higher spin, the so-called Higuchi bound found in Ref. \cite{hig} for massive graviton.  An extensive study of the role of spinning states in de Sitter  has recently been carried out in Refs. \cite{c16,bau} where it was shown that   peculiar  signatures  are present  in the squeezed limit of the correlation functions of the primordial fluctuations with their angular dependence providing  information about the spin. In particular, the link between the Higuchi bound and  the properties of  CFT$_3$, its relation to the (violation of the) tensor consistency relation and the absence of curly hair in de Sitter have been nicely discussed in Ref. \cite{c19}.

The goal of this paper is to reinterpret and discuss   the Higuchi bound, the partial massless states in de Sitter as well as the asymptotic symmetries and  inflationary consistency relations through a  new perspective offered by   radially quantized  CFT$_3$     and the fact that conformal operators are in one-to-one correspondence with the states of the CFT$_3$. We will also discuss the role of symmetries in determining the properties of the correlators in anisotropic inflation.

We will show that the Higuchi bound is a simple consequence of  reflection positivity in radially quantized CFT$_3$
assuring  that
the ground state has finite energy. Indeed, energies must be bounded from below and 
in radial quantization energies of states are the scaling dimensions of operators.  

We will also elucidate  the phenomenon of partial masslessness  on the boundary CFT$_3$ side. Helicities of states of spinning states may become massless in de Sitter  for some particular values
of masses. In the dS/CFT$_3$ correspondence, such  bulk fields correspond to   boundary fields  and through 
the operator-state correspondence of CFT$_3$,  they  correspond on the boundary to  rank-$s$ symmetric
 tensors which are partially conserved.   
 This in turn means that  they  correspond to states which must be    highest weight tensors for the
conformal group: the descendants of the corresponding  symmetric traceless tensors have vanishing norm. 

We will  use the techniques of radial quantization to understand in more detail the role of asymptotic symmetries  and consistency relations
 during inflation. For instance,   asymptotic dilation symmetries   transform asymptotic vacua to new physically inequivalent vacua by generating a long mode for the curvature perturbation. We will show that  on the 
CFT$_3$ side, this corresponds to   evolve  states  forward (or backward) in ``time" along a cylinder obtained by  relating the metric on the Euclidean flat space  ${\mathbb R}^3$ to the metric of a cylinder 
S$^{2}\otimes {\mathbb R}$ by a Weyl tranformation. The 
charge generating the symmetry transformation will be    the Hamiltonian obtained in  the radial quantization  on the cylinder of the CFT$_{3}$, thus allowing a nice interpretation of the asymptotic symmetries in terms of a quantum mechanics analogy. More generically, the 
asymptotic symmetries are generated by the topological charges of the CFT$_3$.

Finally, we will investigate the case of anisotropic de Sitter expansion which is relevant for the case of anisotropic inflation. While anisotropic de Sitter space times are not maximally symmetric, they nevertheless maintain the dilation isometry. This residual symmetry allows to fix the correlators of free four-dimensional scalar fields together with the anisotropy in the power spectrum. By conjecturing that anisotropic de Sitter is dual to a three-dimensional isotropic three-dimensional boundary where the stress energy momentum tensor acquires a nonvanishing 
vacuum expectation value, we are able to reproduce the four-dimensional results.

The paper is organized as follows. In section 2 we provide a summary of the symmetries of de Sitter. Section 3 contains a brief introduction to the notion of radial quantization of CFT$_3$. The Higuchi bound is obtained in section 4, while the case of partially massless states is presented in section 5. Consistency relations and asymptotic symmetries are discussed in section 6. Section 7 contains our findings regarding anisotropic inflation. Conclusions are presented in section 8. The paper contains  also three appendices.

\section{Symmetries of the de Sitter geometry}
\noindent
Let us  summarize here the more salient features about the symmetries of four-dimensional de Sitter. The expert reader can skip this part.

The four-dimensional de Sitter spacetime of radius $H^{-1}$, where $H$ is the Hubble rate can be  described by the hyperboloid \cite{book}
\beq
\eta_{AB}X^AX^B=-X_0^2+\vec{X}^2+X_5^2=\frac{1}{H^2} ~~~A,B=(0,i,,5)\,\,{\rm with}\,\, i=1,2,3, \label{hyper}
\eeq
embedded in  five-dimensional Minkowski spacetime $\mathbb{M}^{1,4}$  with coordinates $X^A$ 
and flat metric  $\eta_{AB}=\rm{diag}(-1,1,1,1,1)$.
A particular parametrization of the de Sitter hyperboloid is then provided by (being $\eta$ the  conformal time)
\be
&&X^0=\frac{1}{2H}\left(H\eta-\frac{1}{H \eta}\right)-\frac{1}{2}\frac{\vx^2}{\eta},\nonumber \\
&&X^i=\frac{x^i}{H\eta},\nonumber \\
&&X^5=-\frac{1}{2H}\left(H\eta+\frac{1}{H \eta}\right)+\frac{1}{2}\frac{\vx^2}{\eta},  \label{poin}
\ee
which satisfies Eq. (\ref{hyper}). The de Sitter metric is the induced metric on the hyperboloid from 
the five-dimensional ambient Minkowski spacetime
\beq
{\rm d} s_5^2=\eta_{AB}\d X^A\d X^B.
\eeq
For the particular parametrization (\ref{poin}) we find the standard de Sitter metric 
\beq
\label{st}
 {\rm d}s^2=g_{\mu\nu}\d x^\mu\d x^\nu=\frac{1}{H^2\eta^2}\left(-{\rm d}\eta^2+{\rm d} \vx^2\right).
\eeq
The group SO(1,4) acts linearly on $\mathbb{M}^{1,4}$. The generators are
\beq
J_{AB}=X_A\frac{\partial}{\partial X^B}-X_B\frac{\partial}{\partial X^A}.
\eeq
We may split them as 
\beq 
\label{JP+}
J_{ij},~~ P_0=J_{05}\, , ~~\Pi^+_i=J_{i5}+J_{0i}\, , ~~\Pi^-_i=J_{i5}-J_{0i}.
\eeq
In this way, they  act on the de Sitter hyperboloid as 
\be
&&J_{ij}=x_i\frac{\partial}{\partial x_j}-x_j\frac{\partial}{\partial x_i},\nonumber \\
&&P_0=\eta\frac{\partial}{\partial\eta}+x_i\frac{\partial}{\partial x_i},\nonumber \\
&& \Pi^-_i=-2H \eta x_i\frac{\partial}{\partial\eta}+H\left(\vx^2\delta_{ij}-2 x_i x_j\right)\frac{\partial}{\partial x_j}
-H \eta^2 \frac{\partial}{\partial x_i},\nonumber \\
&&\Pi^+_i=\frac{1}{H}\frac{\partial}{\partial x_i},
\ee
with  the corresponding commutator relations
\be
\label{dsit}
&&[J_{ij},J_{kl}]=\delta_{il} J_{jk}-\delta_{ik}J_{jl}+\delta_{jk}J_{il}-\delta_{jl}J_{ik},\nonumber \\
&&[J_{ij},\Pi^{\pm}_k]=\delta_{ik} \Pi^{\pm}_{j}-\delta_{jk}\Pi^{\pm}_i,\nonumber\\
&&[\Pi^{\pm}_k,P_0]=\mp \Pi^{\pm}_k, \nonumber\\
&&[\Pi^{-}_i,\Pi^{+}_j]=2J_{ij}+2\delta_{ij}P_0.
\ee
The key point is that this reproduces the  conformal algebra. Indeed, 
\beq
\label{LP}
L_{ij}=iJ_{ij}\, , ~~~D=-i P_0\, , ~~~P_i=-i\Pi^+_i\, , ~~~K_i=i\Pi^-_i, 
\eeq
one has
\be
&&P_i=-\frac{i}{H}\partial_i, \nonumber\\ 
&&D=-i\left(\eta\frac{\partial}{\partial\eta}+x^i\partial_i\right), \nonumber \\
&&K_i=-2iHx_i\left(\eta\frac{\partial}{\partial\eta}+x^j\partial_j\right)-iH(-\eta^2+\vx^2)\partial_i, \nonumber\\
&&L_{ij}=i\left(x_i\frac{\partial}{\partial x_j}-x_j\frac{\partial}{\partial x_i}\right) \label{jpkd},
\ee
 are the  Killing vectors of de Sitter spacetime corresponding to symmetries under space translations $P_i$, dilitations 
$D$, special conformal transformations $K_i$ and space rotations $L_{ij}$. The corresponding  
 conformal algebra has the following commutation rules
\be
\label{conf}
&&[D,P_i]=i P_i \label{DP}, \\
&&[D,K_i]=-iK_i \label{DK}, \\
&&[K_i,P_j]=2i\Big{(}\delta_{ij}D-L_{ij}\Big{)}, \\
&&[L_{ij},P_k]=i\Big{(}\delta_{jk}P_i-\delta_{ik}P_j\Big{)}, \\
&&[L_{ij},K_k]=i\Big{(}\delta_{jk}K_i-\delta_{ik}K_j\Big{)}, \\
&&[L_{ij},D]=0, \label{LD}\\
&&[L_{ij},L_{kl}]=i\Big{(}\delta_{il} L_{jk}-\delta_{ik}L_{jl}+\delta_{jk}L_{il}-\delta_{jl}L_{ik}\Big{)\label{LL}}.
\ee
Let us now take the super-Hubble limit  case $H\eta\ll 1$.  The parametrization (\ref{poin}) reduces to
\be
&&X^0=-\frac{1}{2H^2\eta}-\frac{1}{2}\frac{\vx^2}{\eta},\nonumber \\
&&X^i=\frac{x^i}{H\eta}\nonumber, \\
&&X^5=-\frac{1}{2H^2\eta}+\frac{1}{2}\frac{\vx^2}{\eta},
\ee
where the  hyperboloid degenerates  to the hypercone
\be
-X_0^2+X_i^2+X_5^2=0  \label{cone}.
\ee
By  identifying the  points $X^A\equiv \lambda X^A$  on the cone,  the conformal group  acts linearly and  induces the  conformal transformations $x_i\to x_i'$ with 
\be
&& x_i'=a_i+M_i^jx_j, \\
&& x_i'=\lambda x_i, \\
&&x_i'=\frac{x_i+b_i\vx^2}{1+2b_ix_i+b^2\vx^2}. \label{specconf}
\ee 
on Euclidean $\mathbb{R}^3$ with coordinates $x_i$.  These transformations are the  translations and rotations (generated by $P_i$ and $L_{ij}$ respectively), dilations
(generated by $D$) and special conformal transformations (generated by $K_i$). They act on  the constant time hypersurfaces  
of de Sitter spacetime.   Finally we note  the special conformal transformations can be written in terms of inversion ${\cal I}$
\beq
x_i\to x_i'=\frac{x_i}{\vx^2}, \label{inv}
\eeq
as  

\beq
{\rm special}\,\,\,{\rm conformal}\,\,\,  {\rm transformations}\sim {\cal I}\circ ({\rm translation})\circ {\cal I}.
\eeq 
This property will be useful in the following.
\subsection{Representations}
The representations of the SO(1,4) algebra can be  constructed by using  the method 
of  induced representations. Let us first investigate  
the stability subgroup at the origin of the coordinates $x_i$, that is  is the group  generated by  $(L_{ij},D,K_i)$. 
From the conformal algebra one can see  that $P_i$ and $K_i$ are  raising and lowering operators for the dilation
operator $D$. Therefore there are    states which will be annihilated by $K_i$ and each  irreducible representation  can be   
 specified by an irreducible representation of the rotational group SO(3) ({\it i.e.} its spin) and a definite conformal dimension 
annihilated by $K_i$. Representations  $\phi_s(0)$ of the stability group at the origin 
with spin $s$ and dimension $\Delta$   are  specified by  the following relations  
\be
&&[L_{ij},\phi_s(0)]=\Sigma^{s}_{ij}\phi_s(0), \nonumber \\
&& [D,\phi_s(0)]=-i \Delta \phi_s(0), \nonumber \\
&&[K_i,\phi_s(0)]=0,  \label{ls}
\ee
where we have indicated by $\Sigma^{s}_{ij}$  the spin-$s$ representation of SO(3). 
The representations $\phi_s(0)$  that satisfy the relations (\ref{ls}) are called primary fields and once  the primary fields have been identified, all other fields, called  
the descendants,  are deduced by taking derivatives
of the primaries $\partial_i\cdots \partial_j \phi_s(0)$.   
Furthermore, for all the generators of the stability subgroup,  collectively indicated by  $J=(L_{ij},D,K_i)$, since 
  $\phi(\vec{x})=e^{i\vec{P}\cdot \vx}\phi(0)e^{-i\vec{P}\cdot \vx}$, we have that 
\beq
[J,\phi(\vec{x})]=e^{i\vec{P}\cdot \vx}[\hat{J},\phi(0)]e^{-i\vec{P}\cdot \vx},
\eeq
with 
\beq
\hat{J}=e^{-i\vec{P}\cdot \vx}Je^{i\vec{P}\cdot \vx}=\sum_n\frac{(-i)^n}{n!}x^{i_1}x^{i_2}\cdots x^{i_n}[P_{i_1}[P_{i_2}\cdots[P_{i_n},J],\cdots ]],
\eeq
and $\phi(0)$  a representation of the stability subgroup. 
For  $J=L_{ij},D$  and $J=K_i$ we therefore have 
\be
&&\hat{L}_{ij}=L_{ij}+x_iP_j-x_j P_i, \\
&&\hat{D}=D+x^i P_i,\\
&&\hat{K}_i=K_i+2(x_i D-x^jL_{ij})+2x_i x^jP_j-\vx^2P_i.
\ee
In particular, for scalar degrees of freedom,  the right-hand side of the first equation  in (\ref{ls}) is zero and 
 
\be
&&i [L_{ij},\phi(\vec{x})]=\left(x_i\partial_j-x_j\partial_i\right)\phi(\vec{x}), \\
&&i[K_i,\phi(\vec{x})]=\left(2\Delta x_i+2x_i x^j \partial_j-\vx^2\partial_i\right)\phi(\vec{x}), \\
&&i[D,\phi(\vec{x})]=\left(x^i\partial_i+\Delta\right)\phi(\vec{x}), \\
&&i[P_i,\phi(\vec{x})]=\partial_i \phi(\vec{x}).
\ee
We recall now that the de Sitter algebra SO(1,4) has two Casimir invariants
\be
&&{\cal{C}}_1=-\frac{1}{2}J_{AB}J^{AB}\, , ~~~\\
&&{\cal{C}}_2 = W_A W^A\, , ~~~~~W^A=\epsilon^{ABCDE}J_{BC} J_{DE}.
\ee
Using Eqs. (\ref{JP+}) and (\ref{LP}), it is easy to show that 
\beq
{\cal{C}}_1=D^2+\frac{1}{2}\{P_i,K_i\}+\frac{1}{2}L_{ij}L^{ij} \label{c1}, 
\eeq
which, using the   the explicit  representation   (\ref{jpkd}), becomes
\beq
H^{-2}{\cal{C}}_1=-\frac{\partial^2}{\partial\eta^2}-\frac{2}{\eta}\frac{\partial}{\partial\eta}+\nabla^2.
\eeq
Eq. (\ref{c1}) then gives a fundamental relation between the mass and the conformal weight. Indeed,  
\beq
[{\cal{C}}_1,\phi(0)]=-\Delta(\Delta-3)\phi(0), 
\eeq
implies

\beq
\frac{m^2}{H^2}=-\Delta(\Delta-3)\label{MH}
\eeq
for a massive scalar field in de Sitter.  Appendix A offers a more intuitive way of getting the same relation.

The method of the induced representations for which we have worked out the case of the   scalar can  be adopted  to include higher-spin fields as well.   
For the case of a higher-spin field described by a symmetric-traceless tensor $\phi_{i_1\cdots i_s}$, we obtain
\be
&&i [L_{ij},\phi_{k_1\cdots k_s}]=\left(x_i\partial_j-x_j\partial_i+i\Sigma_{ij}^{s}\right)\phi_{k_1\cdots k_s}, \\
&&i[K_i,\phi_{k_1\cdots k_s}]=\left(2\Delta x_i+2x_i x^j \partial_j-\vx^2\partial_i+2i x^j\Sigma_{ji}^{(s)}\right)\phi_{k_1\cdots k_s}, \\
&&i[D,\phi_{k_1\cdots k_s}]=\left(x^i\partial_i+\Delta\right)\phi_{k_1\cdots k_s}, \\
&&i[P_i,\phi_{k_1\cdots k_s}]=\partial_i \phi_{k_1\cdots k_s},
\ee
and  the spin operator $\Sigma_{ij}^{s}$ acts as 
\beq
\Sigma_{ij}^{s}\phi_{k_1\cdots k_s}=\sum_{\{a\}} 
(\phi_{k_1\cdots k_{a-1}ik_{a+1}\cdots k_s}\delta_{jk_a}-\phi_{k_1\cdots k_{a-1}jk_{a+1}\cdots k_s}\delta_{ik_a}). \label{sigma}
\eeq
It is then easy to verify that 
\beq
{\cal C}_1=-\Delta (\Delta-3)-s(s+1)\,\,\,\,{\rm since }\,\, \,\,\frac{1}{2}\Sigma^{s}_{ij}\Sigma^{s}_{ij}=s(s+1). \label{cas}
\eeq
The equation of motion for the spin $s$ field $\Phi_{\mu_1\cdots \mu_s}$  is 
\begin{eqnarray}
\bigg(\nabla^\mu \nabla_\mu+(s^2-2 s-2)H^2-m^2\bigg)\Phi_{\mu_1\cdots \mu_s}=0, \label{ss}
\end{eqnarray}
where  $\nabla_\mu $ is the covariant derivative in de Sitter space and $\Phi_{\mu_1\cdots \mu_s}$ is subject to the conditions
\begin{eqnarray}
\nabla^{{\mu_1}}\Phi_{\mu_1\mu_2\cdots \mu_s}&=&0,
 ~~~~~(s\geq 1)\nonumber \\
{\Phi^{{\mu_1}}}_{\mu_1\mu_2\cdots \mu_s}&=&0, ~~~~~(s\geq 2). 
\end{eqnarray}
For $m=0$,  Eq. (\ref{ss}) enjoys the extra   gauge invariance transformation
\begin{eqnarray}
\Phi_{\mu_1\cdots \mu_s}\to \Phi_{\mu_1\cdots \mu_s}+\nabla_{(\mu_1}\xi_{\mu_2\cdots \mu_s)}, ~~~{\xi^{\mu_1}}_{\mu_1\cdots \mu_{s-1}}=0 .
\end{eqnarray}
In addition, it can be verified that  \cite{PS}
\begin{eqnarray}
\nabla^\mu \nabla_\mu\Phi_{\mu_1\cdots \mu_s}=\frac{{\cal C}_1+s(s+2)}{H^2}\,  \Phi_{\mu_1\cdots \mu_s},  \label{cas2}
\end{eqnarray}
and therefore, we find from Eqs. (\ref{cas}),  ({\ref{ss}) and (\ref{cas2}) that the mass of the spin $s$ field $\Phi_{\mu_1\cdots \mu_s}$ in de Sitter space is
\beq
\frac{m^2}{H^2}=-\Delta(\Delta-3) +(s-2)(s+1). \label{MHG}
\eeq
Note that  for $s=0$, Eq. (\ref{MHG}) does not coincide with Eq. (\ref{MH}). The reason is that Eq. (\ref{MHG}) for $s=0$ gives the mass of a conformally coupled scalar, that is a scalar whose action is 
\beq
S=\int\d^3x\d\eta\sqrt{-g}\,\left(-\frac{1}{2}(\partial\phi)^2-\frac{1}{12} R\phi^2-\frac{1}{2}m^2\phi^2\right),\label{acts1}
\eeq
where $R=12 H^2$ is the scalar curvature of the de Sitter space. 
%

\section{The radial quantization of CFT$_3$}
This section contains known results too and more details can be found, for instance, in Refs. \cite{ryepfl,tasi}.
In the radial quantization of the CFT$_3$  one foliates the space ${\mathbb R}^3$ by spheres S$^{2}$ centered at the origin, see Fig. 1. The unitary operator that takes points from one sphere to the other is constructed through the dilation operator $D$ as 

\beq
U=e^{iD\ln r_2/r_1}.
\eeq
\begin{figure}[h!]
\vskip -4cm
    \begin{center}
      \includegraphics[scale=.4]{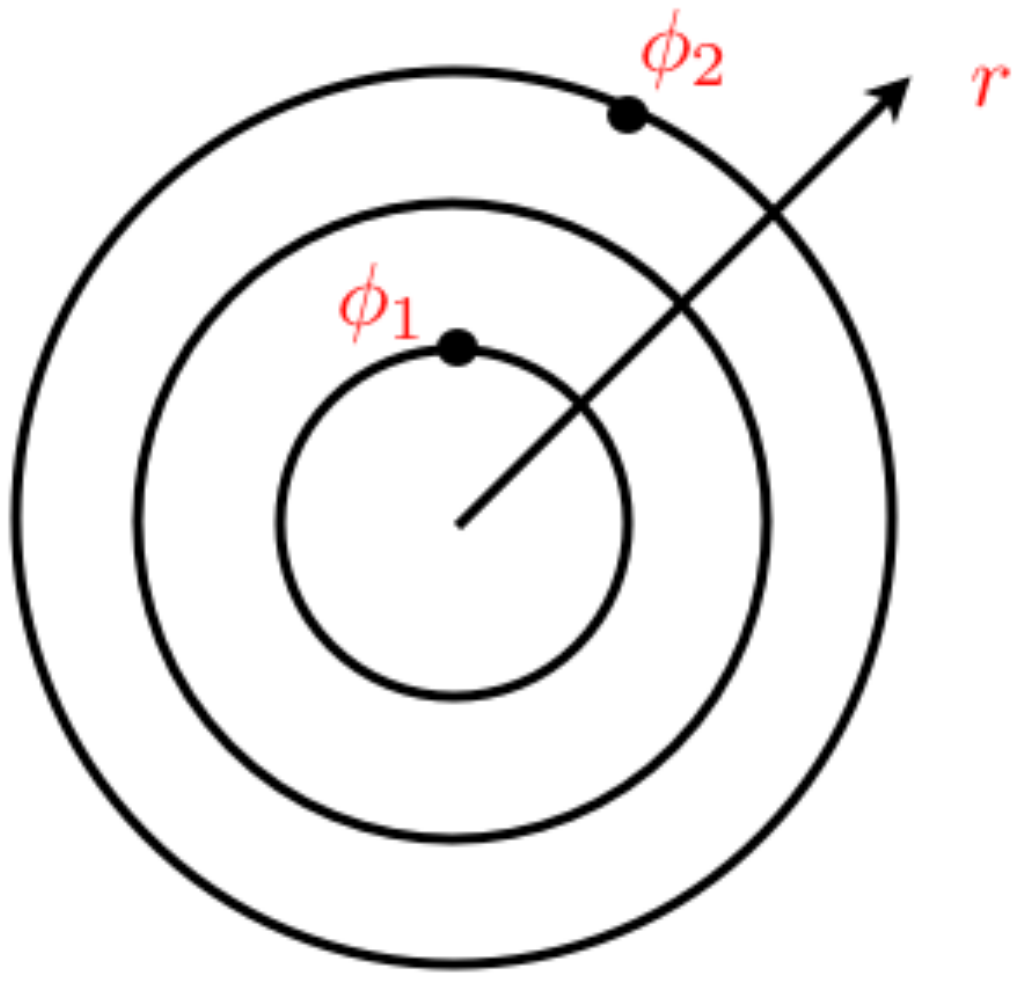}
    \end{center}
     \vskip -4cm
     \caption{\footnotesize A schematic view of the  foliation   of ${\mathbb R}^3$ by spheres S$^{2}$ centered at the origin.}  
\end{figure}
From this expression, one already hints that the operator $D$ plays the role of the Hamiltonian. The  states living on the various spheres are classified according to their scaling dimension $\Delta$

\beq
D|\D\rangle =-i \D |\D\rangle,
\eeq
and possibly by their spin $s$ under the representation of SO($3$) 

\beq
L_{ij}|\D,s\rangle_l=\left(\Sigma_{ij}\right)_l^{\,\,\,l'}|\D,s\rangle_{l'},
\eeq
where we have exploited the fact that the angular momentum $L_{ij}$ is the only one commuting with the operator $D$.
States are therefore  generated by inserting operators inside the sphere. 
Conformal operators are in one-to-one correspondence with the states of the
CFT$_3$

\beq
\lim_{\vx\to 0} \calo_\D(\vx)|0\rangle=\calo_\D(0)|0\rangle=|\Delta\rangle.
\eeq
This is the so-called operator-state correspondence. Thus, by inserting an operator ${\cal O}_\D$ of scaling dimension $\D$ at the origin the state $|\D\rangle={\cal O}_\D(0)|0\rangle$ has also dimension $\D$

\beq
D|\D\rangle=D{\cal O}_\D(0)|0\rangle=\left[D,{\cal O}_\D(0)\right]|0\rangle+{\cal O}_\D(0) D|0\rangle=-i\D{\cal O}_\D(0)|0\rangle=-i\D|\D\rangle.
\eeq
Primary operators are those for which $\vec K|\Delta\rangle=0$ and a conformal multiplet in radial quantization is given by acting with momentum generators on a primary state $|\Delta\rangle$, $P_i|\Delta\rangle$,  $P_iP_j|\Delta\rangle$, $\cdots$. Tthis is equivalent to act with derivatives at the origin

\beq
\lim_{\vx\to 0}\partial_i \calo_\D(\vx)|0\rangle=\left[P_i,\calo_\D(0)\right]|0\rangle=P_i |\Delta\rangle.
\eeq
Since  as the dilation operator moves  points long the spheres, if we insert an operator not at the origin, the corresponding state $|\Psi\rangle={\cal O}_\D(\vx)|0\rangle$ is not an eigenstate of the dilation operator
$D$. We can though decompose the state $|\Psi\rangle$ in terms of 
states with different energies

\beq
|\Psi\rangle={\cal O}_\D(\vx)|0\rangle=e^{i\vec{P}\cdot \vx}{\cal O}_\D(0)e^{-i\vec{P}\cdot \vx}|0\rangle=e^{i\vec{P}\cdot \vx}|\D\rangle=\sum_n\frac{1}{n!}\left(i \vec{P}\cdot \vx\right)^n|\D\rangle.
\eeq
Notice that the operator $\vec{P}$, when applied to the state $|\D\rangle$ $n$ times, it  raises the energy  from $\D$ to $(\D+n)$. This is the consequence of the
commutation relation (\ref{DP}). Similarly,  the operator $\vec{K}$, when applied to the state $|\D\rangle$ $n$ times, it  lowers the energy from $\D$ to $(\D-n)$ as a consequence of the
commutation relation (\ref{DK}). We will now discuss the radial quantization mapping the theory on the cylinder.

\subsection{The mapping to the cylinder}
In a conformal field theory we can relate the metric on the Euclidean flat space  ${\mathbb R}^3$ to the metric of a cylinder 
S$^{2}\otimes {\mathbb R}$ by a Weyl tranformation

\beq
\d s^2_{{\mathbb R}^3}=\d r^2+r^2\d \vec{n}^2=r^2\,\d s^2_{\rm cyl}= e^{2\tau}\,\left(\d\tau^2+\d\vec{n}^2\right),
\eeq
where we have introduced the radial coordinates $r$ and $\vec{n}\in$ S$^{2}$ on ${\mathbb R}^3$ and 

\beq
\tau=\ln r.
\eeq
For instance, expressed in these new coordinates on the cylinder, see Fig. 2, the local operator correlation functions of a scalar field with dimension $\Delta$ become

\beq
\langle \phi(r_1,\vec{n}_1)\phi(r_2,\vec{n}_2)\cdots\rangle=\frac{1}{r_1^{\Delta_\phi}}\frac{1}{r_2^{\Delta_\phi}}\cdots f\left(\tau_i-\tau_j,n_i\right).
\eeq
The function $f$ can only depend on differences of the type $(\tau_i-\tau_j)$ and the unit vectors $n_i$ as the scaling factors $r_i^{-\Delta_\phi}$ are already accounted for.
What remains is already scale invariant and therefore can only depend on ratios of the distances $r_i/r_j$ and therefore on the differences  $(\tau_i-\tau_j)$.

\begin{figure}[h!]
    \begin{center}
      \includegraphics[scale=.4,angle=270]{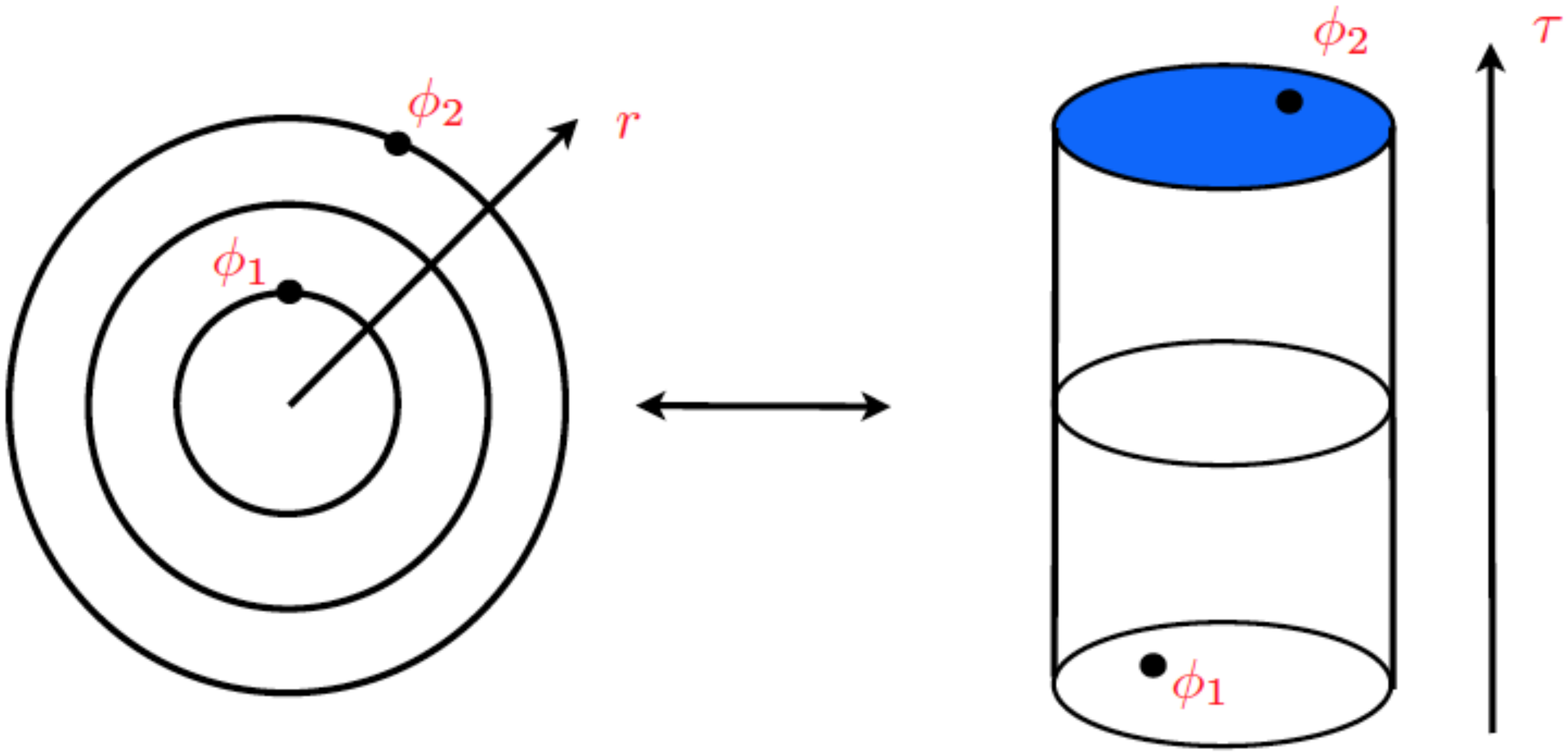}
    \end{center}
    \vskip -2.5cm
     \caption{\footnotesize The map between  ${\mathbb R}^3$ and the cylinder.}  
\end{figure}
 \noindent
This logic suggests a definition of the fields on the cylinder of the type

\beq
\phi_{\rm cyl}(\tau,\vec{n})=r^{\Delta_\phi}\,\phi(r,\vec{n}),
\eeq
where $\phi(r,\vec{n})$ are the fields in ${\mathbb R}^3$. The correlation functions of the fields on the cylinder reduce to

\beq
\langle \phi_{\rm cyl}(\tau_1,\vec{n}_1)\phi_{\rm cyl}(\tau_2,\vec{n}_2)\cdots\rangle=f\left(\tau_i-\tau_j,\vec{n}_i\right),
\eeq
that is the dynamics on S$^{2}\otimes {\mathbb R}$ is invariant under translations. 

On the cylinder we can  define the ``time" reflection operation  which takes $\tau$ into $-\tau$. In the Hamiltonian formulation we have 

\beq
\phi_{\rm cyl}(\tau,\vec{n})=e^{\tau \mathscr{H}_{\rm cyl}}\phi_{\rm cyl}(0,\vec{n})e^{-\tau \mathscr{H}_{\rm cyl}},
\eeq
so that

\beq
\phi^\dagger_{\rm cyl}(\tau,\vec{n})=\left[e^{\tau \mathscr{H}_{\rm cyl}}\phi_{\rm cyl}(0,\vec{n})e^{-\tau \mathscr{H}_{\rm cyl}}\right]^\dagger=e^{-\tau \mathscr{H}_{\rm cyl}}\phi_{\rm cyl}(0,\vec{n})e^{\tau \mathscr{H}_{\rm cyl}}=\phi_{\rm cyl}(-\tau,\vec{n}).
\eeq
This has several consequences:

\begin{enumerate}

\item Any $2n$-point correlatator with operator inserted symmetrically under reflection is  positive in a unitary theory. For instance

\beq
\langle \phi_{\rm cyl}(-\tau,\vec{n})\phi_{\rm cyl}(\tau,\vec{n})\rangle=\langle \phi_{\rm cyl}^\dagger(\tau,\vec{n})\phi_{\rm cyl}(\tau,\vec{n})\rangle\geq 0.
\eeq
This property is called reflection positivity.

\item If we remember that the operators $\vec K$ and $\vec P$ are related by the inversion operator

\beq
\vec K={\cal I} \circ\vec {P}\circ{\cal I},
\eeq
then we have

\beq
\label{a}
\left(\vec{P}|\Psi\rangle\right)^\dagger=\langle\Psi|\vec K\Rightarrow \vec{P}=\vec{K}^\dagger\,\,\,({\rm valid}\,\,\,{\rm only}\,\,\, {\rm in}\,\,{\rm radial}\,\,\,{\rm quantization}).
\eeq
\item Since  ``time" reflection  operation  on the cylinder  corresponds to an inversion ${\cal I}$ in flat space where 
$x_i$ is mapped into $x_i/\vx^2$, this implies that  
 in radial quantization the conjugate of  a state $|\Delta_\phi\rangle=\phi(\vx)|0\rangle$ is given by

\beq
\langle\Delta_\phi|=\langle 0|\phi^\dagger(\vx)\,\,\,\,{\rm with}\,\,\,\, \phi^\dagger(\vx)=r^{-2\D_\phi}\phi({\cal I}\vx)\equiv {\cal I}\left[\phi(\vx)\right].
\eeq
This means that, while the state $|\Delta_\phi\rangle =\lim_{\vec x\to 0}\phi(\vec{x})|0\rangle$ is obtained by acting with $\phi(0)$ on the in vacuum  the state $\langle\Delta_\phi|$ is obtained by acting with $\phi(\infty)$ on the out vacuum:

\beq
\phi(\vx)|0\rangle=e^{i\vec{P}\cdot \vx}\phi(0)e^{-i\vec{P}\cdot \vx}|0\rangle=e^{i\vec{P}\cdot \vx}|\Delta_\phi\rangle,
\eeq
and

\beq
\phi^\dagger(\vx)=e^{i\vec{P}^\dagger\cdot \vx}\phi(\infty)e^{-i\vec{P}^\dagger\cdot \vx}=e^{i\vec{K}\cdot \vx}\phi(\infty)e^{-i\vec{K}^\cdot \vx}\Rightarrow
\langle 0|\phi^\dagger(\vx)=\langle\Delta_\phi|e^{-i\vec{K}^\cdot \vx}.
\eeq
The definition of $\phi(\infty)$ as the conjugate of $\phi(0)$ contains as usual a rescaling factor necessary to get a finite limit, that is 

\beq
\phi(\infty)=\lim_{r\to\infty}
r^{2\Delta_\phi}\phi(\vx).
\eeq
Notice also that, using the expression for the two-point correlator

\beq
\langle 0|\phi(\vx)|\Delta_\phi\rangle=\langle 0| \phi(\vx)\phi(0)|0\rangle=\frac{1}{r^{2\Delta_\phi}},
\eeq
one automatically gets
\beq
 \langle\Delta_\phi|\Delta_\phi\rangle=\lim_{\vx\to 0}r^{-2\D_\phi}\langle 0|\phi({\cal I}\vx)\phi(0)|0\rangle=
\lim_{\vx\to 0}\frac{r^{-2\D_\phi}}{r^{-2\Delta_\phi}}=1.
\eeq

\item If we think as $\tau$ as the ``time" coordinate, in the Schr\"{o}dinger picture, states on the cylinder evolve as

\beq
\partial_\tau |\Psi(\tau)\rangle=-\mathscr{H}_{\rm cyl}|\Psi(\tau)\rangle.
\eeq
A time translation $\tau\to\tau+\lambda$ on the cylinder generates a rescaling $r\to e^\lambda r$. The dilation operator displaces points along the time direction on the cylinder

\beq
D=-i r\partial_r=-i\partial_\tau\Rightarrow \mathscr{H}_{\rm cyl}=iD.
\eeq
In other words,  in radial quantization, states live on spheres, and we evolve from one state to another with the dilation operator. Viewed from the point of view of  the cylinder,  the dilation operator  moves states along the ``time"-direction.
\end{enumerate}
%
%
%
%
%
%
%
%
%
%

\section{The Higuchi bound from the dS/CFT$_3$ correspondence and radial quantization}
The goal of this section is to show that one can deduce  the    Higuchi bound \cite{hig}, which makes impossible  the existence 
spin-1 fields  in de Sitter  spacetime  with  masses $m^2 < 0$ and  
of spin-2 fields with  masses $m^2 < 2H^2$, 
 based on the dS/CFT$_3$ correspondence. 
 
The origin of the  Higuchi bound from a CFT$_3$ point of view has been already nicely discussed in the literature recently
  \cite{c16,c19}. We offer here a  different perspective  using the  radial quantization of CFT$_3$. Essentially, the  Higuchi bound derives from the fact that, even if  primary field satisfies the
 reflection positivity condition on the CFT$_3$ side, a descendant can violate it. This leads to a operator dimension requirement to 
avoid negative norm of the descendants. 

\subsection{First derivation}
Let us  first consider a  spin-1 state in de Sitter. Such a state contains a transverse free and traceless helicity-1 state and 
a helicity-0 state $\pi(\vx)$ of scaling dimension $\Delta_\pi$. According to the dS/CFT$_3$ correspondence, this bulk field corresponds to a dual boundary field $L_\pi(\vx)$ of scaling dimension $h_\pi=3-\Delta_\pi$ and, through the operator-state correspondence, to a state $|h_\pi\rangle$.
The  two-point correlator of the field boundary field $L_\pi$ 
can be written as 

\beq
\langle 0|L_\pi(\vx)L_\pi(0)|0\rangle=\langle 0|L_\pi(\vx)|h_\pi\rangle=\frac{c_\pi}{r^{2 h_\pi}}\Rightarrow 
 \langle h_\pi| h_\pi\rangle=\lim_{r\to \infty} r^{2 h_\pi}\langle 0|L_\pi(\vx)L_\pi(0)|0\rangle=
c_\pi\lim_{r\to \infty}\frac{r^{2 h_\pi}}{r^{2 h_\pi}}=c_\pi> 0,
\eeq
where the last inequality comes from reflection positivity.  On the cylinder we can write

\beq 
P_i=-ie^{-\tau}\left[n_i\partial_\tau+\left(\delta_{ij}-n_i n_j\right)\partial_{n_j}\right].
\eeq
From this expression we deduce the general rules

\begin{eqnarray}
P_i \,e^{-2\tau(h-1)}&=&(-i)n_i(-2)(h-1)e^{-\tau(2 h-1)},\nonumber\\
P_i P_j \,e^{-2\tau(h-1)}&=&\left[-2(h-1)(2h-1)n_i n_j +2(h-1)\left(\delta_{ij}-n_i n_j\right)\right]e^{-2\tau h},\nonumber\\
\vec{P}^2 \,e^{-2\tau(h-1)}&=&\left[-2(h-1)(2h-1)+4(h-1)\right] \,e^{-2\tau h}=2(h-1)(3-2 h)\,e^{-2\tau h}.
\end{eqnarray}
They imply

\begin{eqnarray}
\frac{1}{2}\frac{P_i P_j}{h-1} \,e^{-2\tau(h-1)}&=&\left[-(2h-1)n_i n_j +\left(\delta_{ij}-n_i n_j\right)\right]e^{-2\tau h}
=\left(\delta_{ij}-2 n_i n_j h\right)e^{-2\tau h},\nonumber\\
\frac{1}{2}\frac{\vec{P}^2}{3-2 h} \,e^{-2\tau(h-1)}&=&(h-1) \,e^{-2\tau h},
\end{eqnarray}
so that

\beq
\frac{1}{2}\left(\frac{P_i P_j}{h-1} +\frac{\vec{P}^2\delta_{ij}}{3-2h}\right)\,e^{-2\tau(h-1)}=h\left(\delta_{ij}-2n_i n_j\right)e^{-2\tau h}.
\eeq
With this expression we can now associate to the vector field in the bulk $A^\mu(\vx,\eta)$ of conformal dimension $\D$ a state on the boundary $|L^i\rangle$ for which the conformal dimension is $h=3-\D$ and

\beq
\langle  L^i|L^j\rangle=\left(\delta^{ij}-2n^i n^j\right)=\lim_{r\to \infty}r^{2 h}\langle 0|L^i(\vx)|L^j\rangle=\lim_{\tau\to \infty}\frac{e^{2\tau h}}{2h}\left(\frac{1}{3-2 h}\delta^{ij}\vec{P}^2+\frac{P^i P^j}{h-1}\right)e^{-2\tau(h-1)}.
\eeq
We consider now the scalar descendent $P_i|L^i\rangle$ of the primary vector state $|L^i\rangle$. Since $P_i|L^i\rangle=-\vec{P}^2|\Delta_\pi\rangle$,  if we impose

\beq
\langle P_j L^j|P_i L^i\rangle> 0, 
\eeq
then we get 

\beq
c_\pi=\frac{1}{2 h}\frac{(2-h)}{(2h-3)(h-1)}> 0\,\,\,\,{\rm and}\,\,\,\, h_\pi=h-1.
\eeq
From reflection positivity of the primary helicity-0 state we therefore get the Higuchi bound $h<2$ or $\Delta> 1$. Since

\beq
\Delta=\frac{3}{2}-\sqrt{\frac{1}{4}-\frac{m^2}{H^2}},
\eeq
the Higuchi bound for vectors leads to  $m^2> 0$.  Making the necessary changes, 
we can repeat similar steps to arrive at the Higuchi bound for the massive spin-2 state $\Phi_{\mu \nu}$. 
By defining 
\beq
{C_n}^{i_1\cdots i_n}(\tau,\vec{n},h)=(n_{i_1}\cdots n_{i_n}) e^{-2\tau(h-n/2)},
\eeq
one can show that 
\begin{eqnarray}
{C_1}^i(\tau,\vec{n},h)
&=&
\frac{i}{2 \, (h-1)} \, P^i \, e^{-2\tau(h-1)}, \nonumber \\
{C_2}^{ij}(\tau,\vec{n},h)
&=&
-\frac{1}{4 \, (h-2) \, (h-1)} \, P^i\,P^j \, e^{-2\tau(h-2)}
-\frac{\delta^{ij}}{2\,(h-1)} \,
e^{-2\tau(h-1)}, \nonumber \\
{C_3}^{ijk}(\tau,\vec{n},h)
&=&
- \frac{i}{8 (h-3)(h-2)(h-1)} \,P^i\,P^j \, P^k e^{-2\tau(h-3)} +
\frac{i}{2 (h-1)}\, \big[ \delta^{ij}{C_1}^k\nonumber\\
&+& \delta^{ik}{C_1}^j + \delta^{jk}{C_1}^i \big](\tau,\vec{n},h-1),
\nonumber \\
{C_4}^{ijkm}(\tau,\vec{n},h)
&=&
-\frac{1}{16(h-4)(h-3)(h-2)(h-1)} \,
P^i \, P^j \, P^k \, P^m \, e^{-2\tau(h-4)} \nonumber \\
&-&
\frac{1}{2(h-1)} \, \big[
\delta^{ij} {C_2}^{km} + \delta^{km} {C_2}^{ij} +
\delta^{ik} {C_2}^{jm} + \delta^{jm} {C_2}^{ik} +
\delta^{im} {C_2}^{jk} + \delta^{jk} {C_2}^{im}
\big](\tau,\vec{n},h-1)
\nonumber \\
&+&
\frac{1}{4(h-2)(h-1)} \, (\delta^{ij}\delta^{km} + \delta^{ik}\delta^{jm}
+ \delta^{im}\delta^{jk})\,e^{-2\tau(h-2)}.
\label{nm4}
\end{eqnarray}
In the dS/CFT$_3$ correspondence, a bulk field  $\Phi_{\mu\nu}(\vx,\eta)$ that behaves near the boundary with a given scaling dimension $\Delta$ corresponds to a boundary field $S^{\mu\nu}(\vx)$ of dimension $h=3-\Delta$ and coupling

\beq
\label{gta}
\int_{\eta=0}\d^3 x \,S^{\mu\nu}\Phi_{\mu\nu}
\eeq
and one can compute 
\begin{eqnarray}
\langle  S^{ij}|S^{k\ell}\rangle&=&\lim_{r\to \infty}r^{2h}\langle 0|S^{ij}(\vx)|S^{k\ell}\rangle=\left(\delta^{ij}-2 n^i n^j\right)\left(\delta^{k\ell}-2 n^k n^\ell\right). \end{eqnarray}
From this expression, knowing that $P_iP_j |S^{ij}\rangle= \vec{P}^4|h_\pi\rangle$, and writing

\begin{eqnarray}
 \lim_{\tau\to \infty}\langle 0|S^{ij}(\vx)|S^{k\ell}\rangle&=&\frac{\delta^{ij}\delta^{k\ell}}{4(h-1)(2h-3)}\frac{\vec{P}^4}{(h-2)(2h-5)}\,e^{-2\tau(h-2)}\nonumber\\
 &-&2\delta^{ij}{C_2}^{ij}(\tau,\vec{n},h+1)-2\delta^{k\ell}{C_2}^{ij}(\tau,\vec{n},h+1)+4{C_4}^{ijk\ell}(\tau,\vec{n},h+2),
 \end{eqnarray}
after a lenghty, but straightforward calculation we find

\beq
\langle  S^{ij}P_i P_j|P_kP_\ell S^{k\ell}\rangle>0\Rightarrow
c_\pi=\frac{4-3h+h^2}{4h(2-h)(h-1)(h+1)(5-2h)(2h-3)}\geq 0\,\,\,\,{\rm and}\,\,\,\,h_\pi=h-2.
\eeq
This leads to $h<2$ or $\Delta> 1$. Since 
\beq
\Delta=\frac{3}{2}-\sqrt{\frac{9}{4}-\frac{m^2}{H^2}},
\eeq
one finally gets

\begin{eqnarray}
m^2> 2H^2.
\end{eqnarray}
In the next subsection we derive  the Higuchi bound  in an alternative, and maybe more physically intuitive, manner.  
\subsection{Second derivation}
\noindent
Let us again first consider the simplest case of vector field $A_\mu$. The condition $\nabla^\mu A_\mu=0$ leads on super-Hubble scales to the condition

\beq
(\Delta-2) A_\eta(\vx)=\partial_i A_i(\vx).
\eeq
On the other side, the  operator-state correspondence leads to

\beq
A_{\eta}(\vx)\mapsto \lim_{\vx\to 0}L^{\eta}(\vx)=|L^{\eta}\rangle\,\,\,{\rm and}
\,\,\,\,A_{i}(\vx)\mapsto \lim_{\vx\to 0}{L}^{i}(\vx)=|{L}^{i}\rangle.
\eeq
If $A_{i}(\vx,\eta)$ has scaling dimension $\Delta$, then in the dS/CFT$_3$ correspondence such a bulk field corresponds to the  boundary field $L^i$ of dimension $h=3-\Delta$.
Correspondingly we obtain the following relation

\beq
|L^{\eta}\rangle=\frac{P_i |L^{i}\rangle}{\left(\Delta-2\right)}.
\eeq
Using the conformal algebra  in Eqs. (\ref{conf})-(\ref{LL}), the fact that in radial quantization $P_i^\dagger=K_i$, the transformations of rank-1 tensors $T^{i}$ under rotations

\beq
L_{rs}| T^{i}\rangle=i\left(\delta_{ri}| T^{s}\rangle
-\delta_{si}| T^{r}\rangle\right),
\eeq
and the fact that 
$|L^{i}\rangle$ is a primary

\beq
K_s| L^i\rangle=0,
\eeq
we can now evaluate the norm of the vector $|L^{\eta}\rangle$

\begin{eqnarray}
\label{qweer}
\left(\Delta-2\right)^2\langle L^\eta|L_\eta\rangle &=&\langle L^{r}|   P_i K_r | L^{i}\rangle\nonumber\\
&=&\langle L^{r}|  [P_i,K_r] | L^i\rangle
\nonumber\\
&=&2\left(2-h\right)\langle {L}^{r}  | L_{r}\rangle.
\end{eqnarray}
Imposing that the norm is positive   
we get  again $\Delta>1$ (or $h<2$), which leads to the Higuchi bound. 

For the spin-2 state the procedure is similar.
The equation of motion for the spin $2$ field $\Phi_{\mu \nu}$ in de Sitter  is 
\begin{eqnarray}
\bigg(\nabla^\sigma \nabla_\sigma-2H^2-m^2\bigg)\Phi_{\mu \nu}=0, \label{sts}
\end{eqnarray}
where  $\Phi_{\mu \nu}$ is subject to the conditions
\begin{eqnarray}
\nabla^{\mu}\Phi_{\mu\nu}&=&0,\nonumber \\
{\Phi^{{\mu}}}_{\mu}&=&0.\label{gf}
\end{eqnarray}
 The equation of motion of the helicity-0 part $\Phi_{\eta\eta}$ on super-Hubble scales is

\begin{eqnarray}
\Phi''_{\eta\eta}-\frac{2}{\eta}\Phi'_{\eta\eta}+\frac{m^2}{H^2\eta^2}\Phi_{\eta\eta}&=&0.
\end{eqnarray}
 From this expression we extract that leading time behaviour 
$\Phi_{\eta\eta}(\vx,\eta)\sim (-\eta)^{\Delta}\Phi_{\eta\eta}(\vx)$. The conditions (\ref{gf}) impose
%
%
%
%
%

%
%

\begin{eqnarray}
\label{nm}
\Phi'_{\eta\eta}-\partial_i\Phi_{i\eta}-\frac{1}{\eta}\Phi_{\eta\eta}-\frac{1}{\eta}\Phi_{ii}&=&0,\nonumber\\
\Phi'_{i\eta}-\partial_j\Phi_{ij}-\frac{2}{\eta}\Phi_{i\eta}&=&0,\nonumber\\
\Phi_{\eta\eta}-\Phi_{ii}&=&0,
\end{eqnarray}
from which we deduce 

\beq
(\D-2)(\D-3)\Phi_{\eta\eta}(\vx)=\partial_i\partial_j \hat{\Phi}_{ij}(\vx),\,\,\,\,\hat{\Phi}_{ij}(\vx,\eta)\sim(-\eta)^{\D-2} \hat{\Phi}_{ij}(\vx),
\eeq
where $\hat{\Phi}_{ij}$ stands for the traceless and transverse-free part of ${\Phi}_{ij}$

\beq
\Phi_{ij}=\hat{\Phi}_{ij}+\frac{1}{3}\delta_{ij}\Phi_{ii}.
\eeq
Conformal operators are in one-to-one correspondence with the states of the conformal field theory,

\beq
\Phi_{\eta\eta}(\vx)\mapsto \lim_{\vx\to 0}S^{\eta\eta}(\vx)=|S^{\eta\eta}\rangle\,\,\,{\rm and}
\,\,\,\,\hat{\Phi}_{ij}(\vx)\mapsto \lim_{\vx\to 0}\hat{S}^{ij}(\vx)=|\hat{S}^{ij}\rangle.
\eeq
Correspondingly we obtain the following relation

\beq
|S^{\eta\eta}\rangle=\frac{P_i P_j|{\hat{S}}^{ij}\rangle}{\left(\Delta-3\right)\left(\Delta-2\right)}.
\eeq
Using again the conformal algebra  in Eqs. (\ref{conf})-(\ref{LL}) and 
the transformations of rank-2 tensors $T^{ij}$ under rotations

\beq
L_{rs}| T^{ij}\rangle=i\left(\delta_{ri}| T^{sj}\rangle+\delta_{rj}| T^{is}\rangle
-\delta_{si}| T^{rj}\rangle-\delta_{sj}|T^{ir}\rangle\right),
\eeq
and the fact that 
$|\hat{S}^{ij}\rangle$ is a primary

\beq
K_s|\hat{S}^{ij}\rangle=0,
\eeq
we can now evaluate the norm of the vector $|S^{\eta\eta}\rangle$

\begin{eqnarray}
\label{qwe}
\left(\Delta-3\right)^2\left(\Delta-2\right)^2\langle S^{\eta\eta}|S^{\eta\eta}\rangle&=&\langle \hat{S}^{rs}| K_r K_s P_i P_j| \hat{S}^{ij}\rangle\nonumber\\
&=&\langle \hat{S}^{rs}| K_r [K_s,P_i] P_j| \hat{S}^{ij}\rangle+
\langle \hat{S}^{rs}| K_r P_i [K_s, P_j]| \hat{S}^{ij}\rangle\nonumber\\
&=&4\left(h-2\right)\langle \hat{S}^{rs}| K_r P_j | \hat{S}_{s}^{\,\,\,\,j}\rangle\nonumber\\
&=&8\left(h-2\right)\left(h-3\right)\langle \hat{S}^{r}_{\,\,\,\,s}  | \hat{S}^{s}_{\,\,\,\,\,r}\rangle\nonumber\\
&=&8\Delta\left(\Delta-1\right)\langle \hat{S}^{r}_{\,\,\,\,s}  | \hat{S}^{s}_{\,\,\,\,\,r}\rangle,
\end{eqnarray}
where we have recalled that   conformal dimension of a spin-2 field 
is defined with half indices up and half down \cite{witten} and therefore    raising up  indices increases the corresponding scaling dimension by a factor $+2$.
Imposing that the norm is positive  ($\Delta>0$ is always satisfied)
we get  again $\Delta>1$ (or $h<2$), which leads to the Higuchi bound. 

One can generalize this result to any spin-$s$ state  by requiring that the descendants of the corresponding $s^{\rm th}$-rank symmetric traceless tensor have positive norm

\begin{eqnarray}
|| S^{\overbrace{{\scriptstyle \eta\cdots\cdots\eta}}^{s{\rm -times}}}\rangle||^2&=&\langle 
\hat{S}^{j_1 j_2\cdots j_s}|K_{j_1}K_{j_2}\cdots K_{j_m} P_{i_1}P_{i_2}\cdots P_{i_m}|\hat{S}^{i_1 i_2\cdots i_s}\rangle\nonumber\\
&\sim&(-1)^s m!(h-(s+1))(h-s)\cdots(h-(2+s-m))||\hat{S}^{i_1 i_2\cdots i_s}\rangle||^2.\nonumber\\
&&
\end{eqnarray}
Taking $m=s$, we get again the  bound $h<2$ or $\Delta>1$ which, from Eq. (\ref{MHG}), delivers

\beq
m^2>s(s-1)H^2.
\eeq 
The origin of the Higuchi bound  can therefore be interpreted  from  the state-operator correspondence in radial quantization in the following way.
On  the CFT$_3$ side of the dS/CFT correspondence the various helicities of the  vector correspond to   states
living on a boundary. If  the Higuchi bound is violated a  descendant of these states acquires a negative norm. The Higuchi bound assures  that
the ground state having the smallest energy has finite energy.

\section{The limit of enhanced symmetry and partial masslessness}
As we have seen in the previous section, the case $\Delta=1$ is special for both spin-1 and spin-2 states.
 For $m^2 = 0$, extra  gauge invariance is acquired for vectors and only the helicity-1 modes are physical. For spin-2 states with squared mass $m^2 = 2H^2$, the field becomes partially massless, and the number of propagating degrees of freedom becomes four.  

While in  Minkowski space  massless fields of spin-$s$ state have
helicities $\pm s$ and  massive ones have all helicities running from 
$-s$ to $s$,  in  de Sitter space-time  
there exist  ``partially massless'' fields \cite{partial} with mass

\beq
m^2=H^2 \left[s(s-1)-n(n+1)\right].
\eeq
Helicities  range from  $-s$ to $s$ with  the helicities $-n,-n+1,\dots n$
 removed  for all  $n\leq s-2$.  
Such   fields  are symmetric tensors $\Phi_{\mu_1\cdots\mu_n}$ and the linear action is invariant under the transformation

\beq
\delta\Phi_{\mu_1\cdots\mu_s}=\nabla_{\mu_1}\cdots \nabla_{\mu_{s-n}}\xi_{\mu_{s-n+1}\cdots \mu_s}+\cdots,
\eeq
where the dots stand for terms obtained both by symmetrizing the indices and adding terms with fewer than $s-n-1$ derivatives. 
To the best of our knowledge,  in the literature there is no proof that the gauge-invariance of the partially massless  field can be maintained exactly beyond linear order (see for instance \cite{dr,fz} for  recent discussions on this matter).  

Vectors in de Sitter do not have partial massless states. However, enhanced symmetry is acquired for $m^2=0$.  Spin-2 states have $n=0$ and there exists one partial massless degree of freedom
with mass $m^2=2 H^2$. In both cases the corresponding scaling dimension is $\Delta=1$.

It is interesting to see what the special case $\Delta=1$ corresponds to on the dual theory in terms of states of  spin 1 and 2.
We  discuss first the case of a vector  $A_\mu$, which is maybe less interesting than the case of the massive spin-2 particle, but is technically less involved.
Let us take it massless to start with.
The action is
\be
S=-\frac{1}{4}\int \d^4x \sqrt{-g}g^{\mu\nu}g^{\sigma\rho}F_{\mu\nu}F_{\sigma\rho},
\ee
where $F_{\mu\nu}=\partial_\mu A_\nu-\partial_\nu A_\mu$. We can work in the gauge $A_0$=0 and let us check that the action is
conformally invariant for the  special value $\Delta=1$ of the conformal weight of the fields $A_i$.
We should first recall that a vector $A_i$ 
with scaling dimension $\Delta$ in $D$-dimensions   transforms 
under the conformal group as $A_i(x)\to A_i'(x')$ where
\be
A_i'(x')=\left|\det \left( \frac{\partial {x'}^j}{\partial {x}^i}\right)\right|^{(1-\Delta)/D}
\frac{\partial x^j}{\partial {x'}^i}A_j(x). \label{con}
\ee
Under dilations we have
\be
A_i'(x')=\lambda^{-\Delta}\,  
A_i(x). \label{con1}
\ee
For inversions one gets

\be
A_i'(x')=|\vx|^{2+2(\Delta-1)}J^{\, j}_i(x) A_j(x),
\ee
where one had made use of $\partial_i=\vec{x}'^2\,J^j_{\, i}\partial'_j$, with $J^j_{\, i}=(\delta^j_{\, i}-2x^i x_j/\vec{x}^2)$ (recall that 
 special conformal transformations are obtained by a subsequent operations of inversion, translation and inversion).

The action for the gauge field expressed for the de Sitter metric in conformal time becomes
\begin{align}
S=-\frac{1}{4}\int \d^3x\d\eta
\Big[(\partial_{\t}A_i)^2-\left(\partial_i A_j-\partial_j A_i\right)^2\Big]. \label{sa3}
\end{align}
The  action  can only be invariant under inversions if 
the dimension of the vector $A_i$ is $\Delta=1$. Let us check it. 
Under inversions, the vector $A_i$ transforms as in Eq. (\ref{con}).
This is a coordinate transformation, but  with the extra term
$J^{(1-\Delta)/D}$ where $J=\Big{|}\det(\partial {x'}^j/\partial {x}^i)\Big{|}$. 
When transforming $f_{ij}=(\partial_i A_j-\partial_j A_i)$, there will be  cross terms of the 
form $A_j\partial_iJ$ which cannot be canceled unless  the 
$J$ factor  is missing and this imposes 
$\Delta=1$. 
Then, by using that 
\be
f'_{ij}=|\vx|^4\, J_{ik}\, J_{jl}\, f_{kl}\, , ~~~~\partial_{\t'} A_i'=|\vx|^4 \,J_{ij}\, \partial_{\t} A_j \label{ftr}
\ee
 and the orthogonality relation $J^i_{\, j} J^{j}_{\,k}=\delta^i_{\,k}$, the action is invariant  under  inversion 
\be
S'&=-\frac{1}{4}\int \d^3x'\d\eta' 
\left\{(\partial_{\t'}A_i')^2-(\partial'_i A'_j-\partial'_j A'_i)^2\right\}=-\frac{1}{4}\int \d^3x\d\eta\frac{|\vx|^8}{|\vx|^{8}}
\left\{(\partial_{\t}A_i)^2-(\partial_i A_j-\partial_j A_i)^2\right\}
=S.\nonumber\\
 \label{sa3'}
\ee
\noindent
Let us now find the result $\Delta=1$ in a more convoluted manner, but useful for what we wish to obtain later on. We follow  Ref. \cite{witten} here. We assume
that the vector $A_\mu$ has a non-vanishing mass $m$ and action

\beq
S=\int\d^4x\sqrt{-g}\Big[-\frac{1}{2}\nabla_\mu A_\nu \nabla^\mu A^\nu+\frac{1}{2}\left(\nabla^\mu A_\mu\right)^2-\frac{1}{2}m^2 A_\mu A^\mu\Big].
\eeq
In the limit of $m=0$  the action acquires an extra gauge invariance

\beq
\delta A_\mu=\nabla_\mu \xi.
\eeq
 Through  the dS/CFT$_3$ correspondence, we know that a massless field of spin $s$ corresponds on the boundary to a rank-$s$ conserved symmetric
 tensor. This means that in the limit $m=0$ the massless vector must corresponds on the boundary to a partially conserved current $L^\mu$. Indeed, since the coupling 
 
 \beq
 \int_{\eta=0}\d^3x\, L^\mu A_\mu
 \eeq
 should be gauge-invariant, the partial conservation must have the form
 
 \beq
 \label{fl}
 \nabla_\mu L^\mu=0. 
 \eeq
 Furthermore, on the boundary $\eta=0$, the vector field behaves like

\beq
A_i(\vx,\eta)\sim (-\eta)^{\Delta}A_i(\vx).
\eeq
In the dS/CFT$_3$ correspondence, such a bulk field corresponds to the  boundary field $L^i$ of dimension $h=3-\Delta$.
Via the operator-state correspondence of CFT, the vector $L^i$ corresponds to a state $|L^i\rangle$ which must be a highest weight vector for the
conformal group. Indeed, Eq. (\ref{fl}) in flat space reduces to $\partial_i L^i=0$. This means that the first descendent of $|L^i \rangle$, that is
$P_i |L^i \rangle$, must  a null vector. This can occur only for a particular conformal
dimension of $L^i$. Since conformal invariance is obtained for $\Delta=1$, one should then recover $h=2$. Let us check that this is the case. From 


\beq
L_{rs}| T^{i}\rangle=i\left(\delta_{ri}| T^{s}\rangle-\delta_{si}| T^{r}\rangle\right)
\eeq
and the fact that 
$|{L}^{i}\rangle$ is a primary

\beq
K_r|L^i \rangle=0,
\eeq
we get
\begin{eqnarray}
|| P_i |L^i \rangle ||^2&=&\langle L^r |K_r P_i |L^i \rangle\nonumber\\
&=&\langle L^r |[K_r,P_i] |L^i \rangle\nonumber\\
&=&(h-2)\langle L^i|L_i \rangle,
\end{eqnarray}
which vanishes for $h=2$ as it should.

Similar to what discussed for the massless vector state,  for the spin-2 case, from general arguments we know that a partially massless field must correspond in the boundary theory to a partially conserved tensor $L^{\mu\nu}$. Indeed, since the
coupling (\ref{gta})
should be invariant under the extra gauge symmetry 

\begin{eqnarray}
\delta\Phi_{\mu\nu}=\nabla_{(\mu} \nabla_{\nu)} \xi,  \label{gt}
\end{eqnarray}
then 

\beq
\nabla_\mu\nabla_\nu  L^{\mu\nu}=0.
\eeq
Through  the operator-state correspondence the field $L^{\mu\nu}$ corresponds to a state $|S^{\mu\nu}\rangle$ which must be a highest weight vector for the conformal group. The partially conservation of  $S^{\mu\nu}$   implies that a certain level descendant of $|\hat{S}^{ij}\rangle$ is a null vector. 
From Eq. (\ref{qwe}) we see that this is achieved for $h=2$ or  $\Delta=1$ which correctly  corresponds to the state $|S^{\eta\eta}\rangle$ being null.

\subsection{Generalization to higher spins}
What described above can be generalized to    any spin-$s$ state  by requiring that the descendants of the corresponding $s^{\rm th}$-rank symmetric traceless tensor have vanishing norm, corresponding to the fact that is $\hat{S}^{i_1 i_2\cdots i_s}$ is  partially conserved \cite{witten}

\begin{eqnarray}
|| P_{i_1}P_{i_2}\cdots P_{i_m}|\hat{S}^{i_1 i_2\cdots i_s}\rangle||^2&=&\langle 
\hat{S}^{j_1 j_2\cdots j_s}|K_{j_1}K_{j_2}\cdots K_{j_s} P_{i_1}P_{i_2}\cdots P_{i_s}|\hat{S}^{i_1 i_2\cdots i_s}\rangle\nonumber\\
&\sim& m!(h-(s+1))(h-s)\cdots(h-(2+s-m))||\hat{S}^{i_1 i_2\cdots i_s}\rangle||^2.\nonumber\\
&&
\label{mji}
\end{eqnarray}
Imposing that this  norm  vanishes for $m = r$,  but is  nonvanishing for $m < r$, so that 

\beq
\partial_{i_1}\cdots \partial_{i_m}\hat{S}^{i_1 i_2\cdots i_s}=0,
\eeq
then $h = 2 + s-r$. Such state  corresponds in the four-dimensional de Sitter space to a partially massless field with a range of helicities missing, depending on $r$. 

Notice that for $s>2$ there is always a partial massless state for which $\Delta=0$. Indeed, being

\beq
\D=\frac{3}{2}- \sqrt{\left(s-\frac{1}{2}\right)^2-\frac{m^2}{H^2}}
\eeq
 the dominant scaling dimension at $\eta=0$, the scaling dimensions
 of the partial massless states  become

\beq
\D=\frac{3}{2}- \sqrt{\left(s-\frac{1}{2}\right)^2-s(s-1)+n(n+1)}=\frac{3}{2}\pm \sqrt{\frac{1}{4}+n(n+1)}.
\eeq
For 

\beq
n(n+1)=2\Rightarrow n=1\Rightarrow m^2=H^2 \left[s(s-1)-2\right]\,\,\,{\rm for}\,\,\, s>2,
\eeq
one gets $\Delta=0$ and since $n\leq s-2$, such state for which $n=1$ always exists. 
For $s=3$,  there are two  partially massless states, one for $n=1$ and $\Delta=0$ and the other for $n=0$ and $\Delta=1$.
They correspond in Eq. (\ref{mji})  to $m=s-n=2$ and  $m=s-n=3$, respectively. For  $m=r=2$, one has $h=2+3-2=3$, corresponding  to $\Delta=0$; for $m=r=3$, one has 
$h=2+3-3=2$, corresponding  to $\Delta=1$. These values correctly reproduce the conformal weights of the partially massless states.

Partially massless states with $\Delta=0$ might be relevant during inflation because these states will not decay on super-Hubble scales. It remains to be seen if these partial massless states survive beyond linear order. 

\section{Consistency relations from the dS/CFT$_3$ correspondence, radial quantization and   asymptotic symmetries}
 Inflationary consistency relations have attracted a lot of attention since,  if violated,  they  would  rule out single-field models of inflation.
There exist both  scalar consistency relations relating the squeezed limit of the $(n+1)$-point correlators  to the
 $n$-point correlators of scalar perturbations and tensor consistency relations involving tensor and scalar modes \cite{cr} (see also \cite{cz}). Recently, it has been shown that  the consistency relations are in close connection with the 
  asymptotic symmetries of  de Sitter space \cite{sloth,c21} since soft degrees of freedom  produced by the expansion of de Sitter can be interpreted  as the Nambu-Goldstone bosons of spontaneously broken asymptotic symmetries of the de Sitter spacetime. 
  
  We will see that the radial quantization
  allows to  identify the charges generating these asymptotic symmetries with the topological charges of the CFT$_3$. For instance, in the case of
 the  scalar consistency relations we will see that the corresponding charge is nothing else than the Hamiltonian and that the action of the charge  simply  evolves the states on the cylinder forward (or backward) in “time”.
 
 Let us see first how the consistency relations arise in the CFT$_3$ using radial quantization.
 
 \subsection{Scalar consistency relations}
 Suppose that  inflation has generated a long mode $\zeta_L$ (with wavelength larger than the Hubble radius) for the comoving curvature perturbation such that the perturbed metric reads

\begin{equation}
\label{s}
{\rm d}s^2=\frac{1}{H^2\eta^2}\left[-{\rm d}\eta^2+e^{2\zeta_L(\vx)}{\rm d}\vec{x}^2\right].
\end{equation}
Under a dilation symmetry $\eta\to\lambda\eta$ and $\vec x\to \vec{x}'=\lambda \vec x$, the long mode transforms non-linearly as a  Nambu-Goldstone mode

\beq
\zeta_L\to \zeta_L-\lambda.
\eeq
The constant zero mode of the curvature perturbation can be removed (or generated) by simply choosing $\lambda=\zeta_L$. 
Furthermore,  one can approximate the effect of such a constant long-wavelength mode on an $n$-point function as a rescaling of the coordinates

\begin{equation}
\langle\zeta(\vec{x}_1)\cdots \zeta(\vec{x}_n)\rangle_{\zeta_L}=\langle\zeta(\vec{x}'_1)\cdots \zeta(\vec{x}'_n)\rangle.
\end{equation}
This argument implies  that in that case the squeezed limit of the $(n + 1)$-point function would be 

\begin{equation}
\label{rt}
\langle\zeta_{\vec q} \zeta_{\vec{k}_1}\cdots \zeta_{\vec{k}_n}\rangle'_{q\to 0}=P_\zeta(q)\left[3(n-1)+\sum_{a=1}^n\vec{k}_a\cdot\vec{\nabla}_{k_a}\right]\langle \zeta_{\vec{k}_1}\cdots \zeta_{\vec{k}_n}\rangle',
\end{equation}
where we have indicated by $P_\zeta$ the power spectrum and primes indicate we have removed $\pi$'s and Dirac delta functions. For $n=2$, the relation above
provides the famous Maldacena's consistency relation for 
the three-point correlator of the comoving curvature perturbation in the squeezed limit stating that its size is proportional  to the deviation of the two-point function from scale invariance  and therefore proportional to the slow-roll parameters.
 
Similar arguments lead to the so called conformal consistency relation \cite{c00} where the long mode of the curvature perturbation in the metric  can be removed not only at the level of the constant zero mode, but also at its first gradient. This is achieved simply by a special conformal transformations

\beq
\label{sc1}
\vec x\to \vec x+\vec{b}\,\vec{x}^2-2\vec{x}(\vec{b}\cdot\vec{x}),
\eeq
which can be neutralized by transforming the long mode $\zeta_L$ as

\beq
\zeta_L\to \zeta_L+2 \vec{b}\cdot\vec{x},
 \eeq
and taking  $\vec b=-1/2\vec{\nabla}\zeta_L$. Consequently, 
the effect  a constant long-wavelength gradient mode acts on the  $n$-point function as a rescaling of the coordinates (\ref{sc1})

\begin{equation}
\langle\zeta(\vec{x}_1)\cdots \zeta(\vec{x}_n)\rangle_{\vec{\nabla}\zeta_L}=\langle\zeta(\vec{x}'_1)\cdots \zeta(\vec{x}'_n)\rangle,
\end{equation}
which in momentum space becomes

\begin{equation}
\langle\zeta_{\vec q} \zeta_{\vec{k}_1}\cdots \zeta_{\vec{k}_n}\rangle'_{q\to 0}=-\frac{1}{2}P_\zeta(q)q^i\sum_{a=1}^n\left(6\vec{\nabla}^i_{k_a}-k^i_a\vec{\nabla}^2_{k_a}+2\vec{k}_a\cdot\vec{\nabla}^i_{k_a}\right)\langle \zeta_{\vec{k}_1}\cdots \zeta_{\vec{k}_n}\rangle'.
\end{equation}

\subsection{Scalar consistency relations from the dS/CFT$_3$ correspondence and radial quantization}
The fact that 
 the de Sitter isometry SO(1,4) group acts  as conformal group CFT$_3$
 when the fluctuations are on super-Hubble scales allows a simple interpretation of the consistency relations in radial quantization on the cylinder.
 
 Let  us make an infinitesimal conformal transformation of the coordinates on the cylinder $y^\mu\to y^\mu+\xi^\mu(y)$, where we identify  $y^0=\tau$ and $y^i$ with the angular coordinates ($i=1,2$). In particular, we consider a sphere 
which encloses all the points at which we wish to evaluate the correlators and  such that the 
 transformation is conformal within the sphere, and the identity  outside it. This gives
rise to an (infinitesimal) discontinuity on the surface of the sphere $\Sigma$, and, at least classically,  to a modification of
the action $S$ according to (after integrating by parts)
 
 \beq
 \delta S=\int_{\Sigma}\d S_\mu\,\xi_\nu(x)\,T^{\mu\nu}(x),
 \eeq
where   the integral is on the area of the surface of the sphere.
 This change is  balanced by the explicit change in the correlation function under the conformal transformation
 
 \beq
\delta \langle\phi(\vx_1)\phi(\vx_2)\cdots \phi(\vx_n)\rangle=-\int_{\Sigma}\d S_\nu\,\xi_\mu(x)\,\langle T^{\mu\nu}(x)\phi(\vx_1)\phi(\vx_2)\cdots \phi(\vx_n)\rangle.
 \eeq
\begin{figure}[h!]
\vskip -3.5cm
    \begin{center}
      \includegraphics[scale=.4]{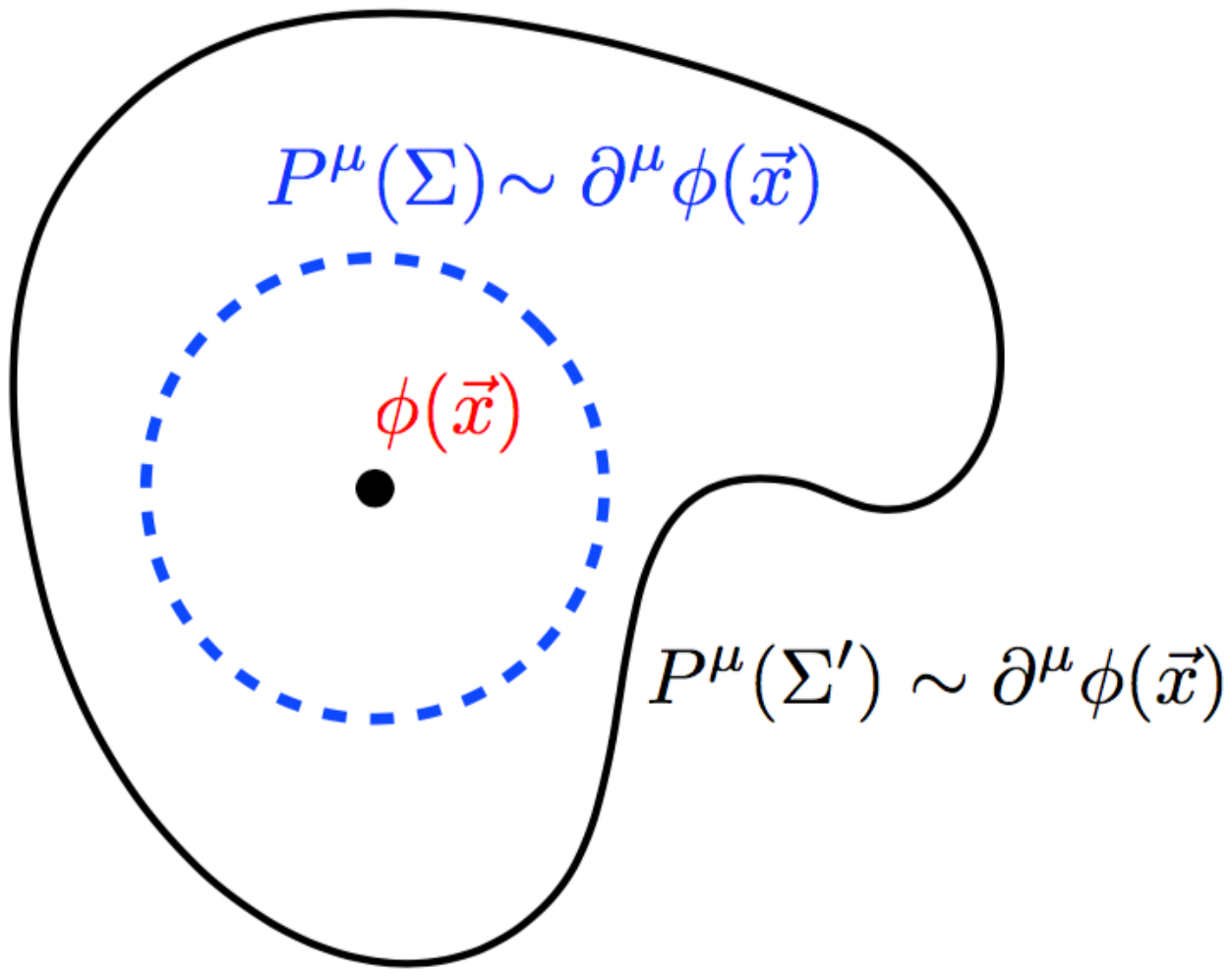}
    \end{center}
     \vskip -4cm
     \caption{\footnotesize A surface $\Sigma$ supporting the operator $P^\mu(\Sigma)$  can be deformed  without changing the correlation function if the deformation does not  cross any operator insertions.}  
\end{figure}
 In particular, if we perform a dilation transformation $r\to e^\lambda r$, this corresponds to a shift in the cylinder coordinate $\tau\to \tau+\lambda$, that is
 $\xi^0=\lambda$ and $\xi^i=0$. Therefore
 
  \beq
\delta_\lambda \langle\phi(\vx_1)\phi(\vx_2)\cdots \phi(\vx_n)\rangle=-\lambda\int_{\Sigma}\d S_\nu\,\langle T^{\nu 0 }(x)\phi(\vx_1)\phi(\vx_2)\cdots \phi(\vx_n)\rangle=\lambda\langle P^0(\Sigma) \phi(\vx_1)\phi(\vx_2)\cdots \phi(\vx_n)\rangle,
\eeq
where in the last passage we have introduced the topological surface operator
\beq
\label{top}
P^\mu(\Sigma)=-\int_{\Sigma}\d S_\nu\,T^{\mu\nu}(\vx),
\eeq
  for which
  
  \beq
  P^0= \mathscr{H}_{\rm cyl} =iD=\partial_\tau.
  \eeq
  Ward identities guarantee that the correlator of $P^\mu(\Sigma)$ with other operators
is unchanged as we move the surface, as long as it does not cross any operator
insertions. Indeed, from the Ward identity

\beq
\partial_\mu  \langle T^{\mu\nu}(\vx)\phi(\vx_1)\phi(\vx_2)\cdots \phi(\vx_n)\rangle=-\sum_{i=1}^n\delta(\vx-\vx_i)\partial_i^\nu \langle\phi(\vx_1)\phi(\vx_2)\cdots \phi(\vx_n)\rangle,
\eeq
 integrating over the  boundary of a ball  containing, say  $\vx_1$  and no other insertions, one concludes that the results are  independent from the surface.

  Since $\langle P^0(\Sigma)\phi(\vx)\cdots\rangle=\partial_\tau\langle\phi(\vx)\cdots\rangle$, identifying the scalar field with the comoving curvature
  perturbation   $\zeta$ and $\lambda$ with its long mode $\zeta_L$, going to Fourier space one finds the expression (\ref{rt}). This operation amounts
  to leaving the decoupling limit in which gravity is not dynamical.  Adding gravity, things may be more transparent in  the $\zeta$-gauge where the inflaton field driving inflation is unperturbed
  and the time slicing is fixed. In such a case,  and since we wish to obtain for the time being only conformal rescaling of the spatial part of the metric,  transformations of the three-dimensional  conformal group SO(1,4) for every fixed time can be performed.  
 When gravity is switched off,   SO(1,4) is  a non-linearly realized symmetry of the action in de Sitter,  
 in the presence of gravity
 SO(1,4) is the symmetry group of  CFT$_3$.

Since the  long wavelength states $\zeta_L$ populate the future boundary of de Sitter where the CFT$_3$ is living, we can interpret these  scalars as  the Nambu-Goldstone bosons of spontaneously broken asymptotic symmetries of the de Sitter spacetime. These scalars are physical    adiabatic modes \cite{weinberg}. The associated  charge  \cite{c21} has therefore a nice interpretation in the framework of radial quantization: it is nothing else that the Hamiltonian 
which in radial quantization is associated with the dilation operator.

Acting with this charge on the state creates a new state equivalent to a change in the local coordinates induced by the soft scalar 
and adding a constant long wavelength mode
results in evolving the states  forward (or backward) in ``time" along the cylinder through the Hamiltonian. This charge operates on  states by evolving them to other states

\beq
|\D_\zeta\rangle\to e^{\zeta_L \mathscr{H}_{\rm cyl} }|\D_\zeta\rangle=e^{i\zeta_L D}|\D_\zeta\rangle=
r^{-\zeta_L\Delta_\zeta}|\D_\zeta\rangle,
\eeq
in a way that correlators feel the evolution only through scaling dimensions.
By expanding in powers of small $\zeta_L$ one finds the standard result that the dependence of the short
modes on the long mode is proportional to the scaling dimension of $\zeta$. This scaling dimension differs from zero only through the slow-roll parameters.

\subsection{Another  perspective}
All  considerations above  indeed follow from the basic  property that correlation functions on conformally flat backgrounds in CFT$_3$ can be calculated by rescaling the flat space correlation functions

 \beq
 \label{yx}
 \langle\phi(\vx)\cdots\rangle_{e^{2\Omega(\vx)}\d\vec{x}^2}=e^{-\Omega(\vx)\Delta}\langle\phi(\vx)\cdots\rangle_{{\mathbb R}^3}.
 \eeq
This on the cylinder becomes

\beq
\phi_{\rm cyc}(\tau,\vec n)=r^{\Delta}\phi(\vx).
\eeq
Let us demonstrate the rule (\ref{yx})  by computing the two point function on the cylinder starting from a simple  dilation $r\to e^\lambda r$. 

This rescaling corresponds to 
 a time translation $\tau\to\tau+\lambda$ on the cylinder  and that the dilation operator displaces points along the ``time" direction on the cylinder. 
Consider the two-point function of the curvature perturbation $\zeta(\vx)$ of scaling dimension $\Delta_\zeta$

\beq
\langle \zeta(r_1,\vec{n}_1)\zeta(r_2,\vec{n}_2)\rangle=\frac{1}{r_1^{\Delta_\zeta}}\frac{1}{r_2^{\Delta_\zeta}}
\langle \zeta_{\rm cyl}(\tau_1,\vec{n}_1)\zeta_{\rm cyl}(\tau_2,\vec{n})\rangle.
\eeq
In the Hamiltonian formulation we can then write

\beq
\langle \zeta(r_1,\vec{n}_1)\zeta(r_2,\vec{n}_2)\rangle=\frac{1}{r_1^{\Delta_\zeta}}\frac{1}{r_2^{\Delta_\zeta}}
\langle e^{\tau_1 \mathscr{H}_{\rm cyl}}\zeta_{\rm cyl}(0,\vec{n}_1)e^{-\tau_1 \mathscr{H}_{\rm cyl}}\,e^{\tau_2 \mathscr{H}_{\rm cyl}}\zeta_{\rm cyl}(0,\vec{n}_2)e^{-\tau_2 \mathscr{H}_{\rm cyl}}\rangle.
\eeq
Since the Hamiltonian is the dilation operator we can also write

\beq
\langle \zeta(r_1,\vec{n}_1)\zeta(r_2,\vec{n}_2)\rangle=\frac{1}{r_1^{\Delta_\zeta}}\frac{1}{r_2^{\Delta_\zeta}}
\langle e^{i\tau_1 D}\zeta_{\rm cyl}(0,\vec{n}_1)e^{-i\tau_1 D}\,e^{i\tau_2 D}\zeta_{\rm cyl}(0,\vec{n}_2)e^{-i\tau_2 D}\rangle.
\eeq
The dilation $r\to r'=e^\lambda r$ does not have an impact on the cylinder (remember that the dilation operator leaves the vacuum invariant) and one can therefore write

\begin{eqnarray}
\langle \zeta(r'_1,\vec{n}'_1)\zeta(r'_2,\vec{n}'_2)\rangle_\lambda&=&
\frac{1}{{r'}_1^{\Delta_\zeta}}\frac{1}{{r'}_2^{\Delta_\zeta}}
\langle e^{i\tau'_1 D}\zeta_{\rm cyl}(0,\vec{n}_1)e^{-i\tau'_1 D}\,e^{i\tau'_2 D}\zeta_{\rm cyl}(0,\vec{n}_2)e^{-i\tau'_2 D}\rangle\nonumber\\
&=&\frac{1}{{r'}_1^{\Delta_\zeta}}\frac{1}{{r'}_2^{\Delta_\zeta}}
\langle e^{i(\tau_1+\lambda) D}\zeta_{\rm cyl}(0,\vec{n}_1)e^{-i(\tau_1+\lambda) D}\,e^{i(\tau_2+\lambda) D}\zeta_{\rm cyl}(0,\vec{n}_2)e^{-i(\tau_2+\lambda) D}\rangle\nonumber\\
&=&e^{-2\lambda\Delta_\zeta}\frac{1}{r_1^{\Delta_\zeta}}\frac{1}{r_2^{\Delta_\zeta}}
\langle \zeta_{\rm cyl}(\tau_1,\vec{n}_1)\zeta_{\rm cyl}(\tau_2,\vec{n})\rangle\nonumber\\
&=&e^{-2\lambda\Delta_\zeta}\langle \zeta(r_1,\vec{n}_1)\zeta(r_2,\vec{n}_2)\rangle,
\end{eqnarray}
which is the relation (\ref{yx}) with $\Omega(\vx)=\lambda$.  Here we have also made use of the basic property that a dilation transformation does not
change  the vectors $\vec{n}_1$ and $\vec{n}_2$ since the two-point correlator depends only on such an angle, we have had the freedom to 
replace the vectors $\vec{n}_i$ with the vectors $\vec{n}_i$. Appendix B offers an alternative way to prove that that correlation functions on conformally flat backgrounds  can be calculated by rescaling the flat space correlation functions.

 Similar considerations hold for the conformal consistency relation in which one removes the constant gradient of $\zeta_L$ for which Eq. (\ref{yx}) holds for $\Omega(\vx)=-2\vec b\cdot\vx=\vec{\nabla}\zeta_L\cdot\vx$. Here one has to  make again use of the basic property that the special conformal transformation   does not
change  the angle between the vectors $\vec{n}_1$ and $\vec{n}_2$ and since the two-point correlator depends only on such an angle, there is the freedom to 
replace the vectors $\vec{n}'_i$ with the vectors $\vec{n}_i$.

\subsection{Tensor consistency relation}
As for the  consistency relations involving tensors, one can generate a long wavelength tensor mode by  the transformations of the coordinates 

\beq
x^i\to x'^{\,i} =x^{i} +\frac{1}{2} \gamma^L_{ij} \,x^j,
\eeq
 where $\gamma_{ij}$ is the traceless transverse tensor mode. Considering the long mode as a background if seen on small scales, one can write the relation

 \begin{equation}
\langle\zeta(\vec{x}_1)\cdots \zeta(\vec{x}_n)\rangle_{\gamma_L}=\langle\zeta(\vec{x}'_1)\cdots \zeta(\vec{x}'_n)\rangle.
\end{equation}
 Expanding at linear order in the long tensor mode and going to momentum space one finds
 
 \beq
 \label{po}
\langle \gamma^s_{\vec{q}}\zeta_{\vec{k}_1}\zeta_{\vec{k}_2}\rangle'_{q\to 0}\simeq\frac{3}{2}\epsilon^s_{ij}\frac{k_1^i}{k_1}\frac{k_1^j}{k_1}P_\gamma(q)
\langle \zeta_{\vec{k}_1}\zeta_{\vec{k}_2}\rangle'+\cdots,
\eeq
where we have introduced the polarization vectors by $\gamma^{ij}_{\vec{q}}=\sum_{s=1,2}\epsilon^s_{ij}\gamma^s_{\vec{q}}$, $P_\gamma$ is the power spectrum of the tensor mode and the dots  stand for  terms sub-leading in $\Delta_\zeta$. 
 Notice that  the tensor consistency relation is then a consequence of the fact that the partition function in the CFT$_3$ side is invariant under diffeomorphism. The  transformation generating the long tensor mode
cannot be reproduced by a conformal transformations since the metric $\delta_{ij}\d x^i\d x^j$ on ${\mathbb R}^3$ transforms into $(\delta_{ij}+\gamma_{ij})
\d x^i\d x^j$ and this is not of the form dictated by conformal transformations $\delta_{ij}\d x^i\d x^j \to e^{2\Omega(\vx)}\delta_{ij}\d x^i\d x^j$. This is the reason    why
the tensor consistency relation is  not suppressed by deviation from de Sitter.

\subsection{Tensor consistency relation from the  dS/CFT$_3$ correspondence and radial quantization}
Let us reproduce the tensor consistency relation starting from the CFT$_3$ and radial quantization. We follow the procedure of the previous subsection, with the
appropriate differences.
%
%
First, we locate the operators
$\zeta(\vx)$ close to the origin and consider the long tensor mode constant on a sphere surrounding such operators.  As we mentioned already, under a general infinitesimal non-conformal coordinate transformation $x^i\to x^i+\xi^i(\vx)$, the action response has the form

\beq
\delta S=-\int\d^3 x\, \partial^i\xi^j(\vx) T_{ij}(\vx),
\eeq
where $T_{ij}$ is the stress-energy momentum tensor.
Since the  correlation functions involving $\zeta(\vx)$ with respect to the original action are equal to those of $\zeta(\vx)+\delta\zeta(\vx)$ with respect to the modified action, one finds close  to the origin and sphere of radius $\epsilon$ 

\beq
\delta\zeta(\vec\epsilon)=\int_{r>\epsilon}\d^3 x\, \partial^i\xi^j(\vx) T_{ij}(\vx)\zeta(0),
\eeq
where the integration is over the complement of  the sphere of radius $\epsilon$ surrounding the operator at the origin. By integrating by parts and using the conservation of energy one finds

\beq
\delta\zeta(\vec\epsilon)=-\int_{r=\epsilon}\d S^i\, \xi_j(\vx) T_{ij}(\vx)\zeta(0)=-\frac{1}{2}\gamma_L^{jk}\epsilon_k \int_{r=\epsilon}\d S^i\,T_{ij}\zeta(0)=
\frac{1}{2}\gamma_L^{jk} \epsilon_k\,P_j(\Sigma_\epsilon)\zeta(0),
\eeq
where $\Sigma_\epsilon$ is the surface surrounding the sphere.   We find
\beq
\delta\zeta(\vx)=
\frac{1}{2}\gamma_L^{jk} x_k\,P_j(\Sigma)\zeta(\vx),
\eeq
where we have used again the topological operator (\ref{top}) and the fact that the result is independent of the surface as long as we do not cross any operator.
Since $\langle P_i(\Sigma)\zeta(\vx)\cdots\rangle=\partial_i\langle\zeta(\vx)\cdots\rangle$ we see that the operator $\zeta(0)$ creating the state
$|\D_\zeta\rangle$ gets shifted by an amount $ 1/2\gamma^{ij}_L x_j\partial_i\zeta(\vx)|0\rangle$ evaluated at the origin.  Going to momentum-space
one thus recovers the consistency relation (\ref{po}). 

In agreement with Ref. \cite{sloth}, we see that soft gravitons produced by the  de Sitter expansion  can be viewed as the Nambu-Goldstone bosons of spontaneously broken asymptotic symmetries of the de Sitter spacetime. The corresponding charge is    the  topological operator 
\beq
Q_\gamma=-\int_{r=\epsilon}\d S^i\, \xi_j(\vx) T_{ij}(\vx),\,\,\,\, \xi_j(\vx)=\frac{1}{2}\gamma^{ij}_L x_j.
\eeq
We conclude that asymptotic symmetries are generated by the topological charges of CFT$_3$.

\section{Anisotropic de Sitter}
In this section we consider the case of anisotropic inflation, see for instance Refs.  \cite{wise,soda1,soda2} and \cite{reviewanis} for a review. The spacetime can be approximated by a de Sitter expansion with different Hubble rates along different directions. Our goal is to obtain some general results based on symmetry arguments and to find a dual interpretation of them. The first step is to study the corresponding  isometries.

\subsection{The isometries of anisotropic de Sitter}
Let us consider a metric of the form (we use cosmic time)

\beq
\d s^2=\d t^2- \sum_{i,j=1}^3 \delta_{ij}\,a_i^2(t)\,\d x^i\d x^j=\d t^2- \sum_{i,j=1}^3\, \delta_{ij}\,e^{2 H_i t}\, \d x^i\d x^j, \label{ds1}
\eeq
parametrizing an anisotropic de Sitter expansion with unequal expansion rates $H_i$ along the three cartesian axes. The  isometries are transformations of the form 

\beq
x^\mu\to x^\mu+\xi^\mu(x)
\eeq
 which leave
the metric invariant. The infinitesimal functions $\xi^\mu(x)$ 
are the solutions of the Killing equation
\be
\nabla^{\mu} \xi^{\nu} + \nabla^{\nu} \xi^{\mu} = 0,
\label{Kil}
\ee
being  $\nabla^{\mu}$ the covariant derivative.
Recasting   Eq. (\ref{Kil}) under the form

\be
g_{\nu\lambda}\partial_\mu \xi^\lambda+g_{\mu\lambda}\partial_\nu \xi^\lambda+\partial_\sigma g_{\mu\nu}\xi^\sigma=0,
\ee
we find the following set of equations

\begin{eqnarray}
\partial_{t}\xi^t &=& 0,\nonumber\\
\partial_i\xi^{t}+\sum_{j=1}^3 \delta_{ij} \,a_j^2 \,\partial_t\xi^j  &=&0,\nonumber\\
\sum_{k=1}^3\left(\delta_{jk}\, a_k^2\,\partial_i\xi^k+\delta_{ik}\, a_k^2\,\partial_j\xi^k +2\delta_{ij}\, \dot{a}_i a_i\,\xi^t\right)&=& 0.
\label{cKv}
\end{eqnarray}
As one can imagine, the isometries are much less rich than in the isotropic de Sitter case.  We loose of course the three-dimensional isotropy and  the special conformal symmetry, but we keep

\begin{enumerate}

\item  three translations with  Killing vectors

\beq
\xi^{t} =0\,\,\,\,\,{\rm and} \,\,\,\,\,\xi^i={\rm constant},
\eeq

\item dilations $\mathscr{D}$ with Killing vectors

 \be
\,\,\,\,\,\,\,\,\,\,\,\,\,\,\, \xi^t = -1\,\,\,\,\,{\rm and} \,\,\,\,\  \xi^i =\sum_{j=1}^3 \delta_{ij}\, H_i\, x^i.
 \ee
These  dilational Killing vectors are the infinitesimal form of the
finite dilational symmetry,

\beq
\label{tranani}
t\to t-H^{-1} \ln\lambda\,\,\,\,\,\,\,{\rm and} \,\,\,\,\,\,\,
x^i \to \lambda^{1+\epsilon_i}\,x^i,
\eeq
where we have defined the average expansion rate $H$ and the corresponding rate deviation $\epsilon_i$ from isotropy

\beq
H=\frac{1}{3}\sum_{i=1}^3 H_i,\,\,\,\,\,\,\,
\epsilon_i=\frac{H_i-H}{H}\,\,\,\,\,\,\,\,\,{\rm with} \,\,\,\,\,\,\,\,\
\sum_{i=1}^3\epsilon_i= 0.\label{prop}
\eeq
\end{enumerate}
Dilations will play a crucial role in what follows. For convenience, we write the  main equations using the conformal time  
\beq
\eta=-\frac{1}{H}e^{-Ht},
\eeq
defined with respect to the isotropic Hubble rate $H$. The metric (\ref{ds1}) becomes

\beq
\d s^2=\frac{1}{H^2\eta^2}\left[\d \eta^2- \sum_{i,j=1}^3 \delta_{ij} (-H\eta)^{-2\epsilon_i}\d x^i\d x^j\right],
\label{metricc}
\eeq
which is invariant under the dilation transformations

\beq
\label{trananii}
\eta\to \lambda\eta\,\,\,\,\,\,\,{\rm and} \,\,\,\,\,\,\,
x^i \to \lambda^{1+\epsilon_i}\,x^i.
\eeq
%

\subsection{Correlators of free fields in anisotropic de Sitter}
We investigate  here the simplest case of a     free massive spectator
field $\sigma(\vec{x},\eta)$. 
The action

\begin{eqnarray}
S=\int \frac{\d \eta \d^3 x}{H^2\eta^2}  \left[\frac{1}{2}\left(\partial_\eta\sigma\right)^2-\frac{1}{2}\sum_{i=1}^3(-H\eta)^{2\epsilon_i}\left(\partial_i\sigma\right)^2-\frac{1}{2}\frac{m^2}{H^2\eta^2}\sigma^2
\right]
\end{eqnarray}
 is manifestly invariant under the transformations (\ref{trananii}) if the scalar field  transforms  as a scalar.
 Let us  reparametrize these  transformations as 
 
 \beq
\eta\to (1+\lambda)\eta,\,\,\,\,\,\,\,\, x_i\to x_i+C_{ij}\,x_j\,\,\,\,\,\,\,\,{\rm with}\,\,\,\,\,\,\,\,C_{ij}=\lambda(1+\epsilon_i)\delta_{ij} \,\,\,\,\,\,\,\,(i=1,2,3)
 \eeq
and let us see what the invariance implies. We write down the $n$-point correlator of the field $\sigma$. The variation of the latter is

\begin{eqnarray}
\delta\langle\sigma(\vx_1,\eta)\cdots\sigma(\vx_n,\eta)\rangle&=& \left( \lambda\eta\partial_\eta +
\sum_{a=1}^nC_{ij} x_{aj}\partial_{ai}\right)\langle\sigma(\vx_1,\eta)\cdots \sigma(\vx_n,\eta)\rangle.
\end{eqnarray}
Going to momentum space we obtain 

\begin{eqnarray}
\delta\langle\sigma(\vx_1,\eta)\cdots\sigma(\vx_n,\eta)\rangle&=&\int\frac{\d^3 k_1}{(2 \pi)3}\cdots\int\frac{\d^3 k_n}{(2 \pi)3}\sum_{a=1}^n \delta^{(3)}(\vk_1+\cdots+\vk_n)\langle\sigma_{\vk_1}\cdots\sigma_{\vk_n}\rangle
\nonumber\\
&\times& C_{ij} k_{ai}\left(\partial_{k_{aj}}\,e^{i\vk_1\cdot \vx_1+\cdots+i\vk_n\cdot \vx_n}\right)\nonumber\\
&+&n\Delta\int\frac{\d^3 k_1}{(2 \pi)3}\cdots\int\frac{\d^3 k_n}{(2 \pi)3}\sum_{a=1}^n \delta^{(3)}(\vk_1+\cdots+\vk_n) \langle\sigma_{\vk_1}\cdots\sigma_{\vk_n}\rangle \,e^{i\vk_1\cdot \vx_1+\cdots+i\vk_n\cdot \vx_n}\nonumber\\
&&
\end{eqnarray}
Integrating by parts and imposing that the variation is vanishing we get

\begin{eqnarray}
0&=&-\int\frac{\d^3 k_1}{(2 \pi)3}\cdots\int\frac{\d^3 k_n}{(2 \pi)3}\sum_{a=1}^n \delta^{(3)}(\vk_1+\cdots+\vk_n)\langle\sigma_{\vk_1}\cdots\sigma_{\vk_n}\rangle
\nonumber\\
&\times& C_{ij} \left(\partial_{k_{aj}}k_{ai}\right)\,e^{i\vk_1\cdot \vx_1+\cdots+i\vk_n\cdot \vx_n}\nonumber\\
&-&\int\frac{\d^3 k_1}{(2 \pi)3}\cdots\int\frac{\d^3 k_n}{(2 \pi)3}\sum_{a=1}^n \delta^{(3)}(\vk_1+\cdots+\vk_n)C_{ij} k_{ai}\nonumber\\
&\times&\left(\partial_{k_{aj}}\langle\sigma_{\vk_1}\cdots\sigma_{\vk_n}\rangle\right)
\,e^{i\vk_1\cdot \vx_1+\cdots+i\vk_n\cdot \vx_n},\nonumber\\
&+&n\Delta\int\frac{\d^3 k_1}{(2 \pi)3}\cdots\int\frac{\d^3 k_n}{(2 \pi)3}\sum_{a=1}^n \delta^{(3)}(\vk_1+\cdots+\vk_n) \langle\sigma_{\vk_1}\cdots\sigma_{\vk_n}\rangle \,e^{i\vk_1\cdot \vx_1+\cdots+i\vk_n\cdot \vx_n},\nonumber\\
\label{long}
&&
\end{eqnarray}
where we have used the fact that $C_{ij} k_{ai}\partial_{k_{aj}}\delta^{(3)}(\vk_1+\cdots+\vk_n)=0$,  the last of Eq. (\ref{prop}) 
and we have introduced the 
 scaling dimension of the scalar field. 
Finally, Eq. (\ref{long}) becomes

\begin{eqnarray}
\Big[3(n-1)-n\Delta +\sum_{a=1}^n (1+\epsilon_i)\vec{k}_{ai}\cdot\vec{\nabla}_{k_{ai}}\Big]\langle\sigma_{\vk_1}\cdots\sigma_{\vk_n}\rangle'=0.
\end{eqnarray}
At this point we should stress that the momenta involved in this relation are those with indices down, which do not depend on time in anisotropic backgrounds.
For the case of the two-point correlator, one  can easily check that the solution is 

\begin{eqnarray}
\label{iu}
\langle\sigma_{\vk}\sigma_{-\vk}\rangle'&=& \frac{H^2}{2k^{3-2\Delta}}\left[1+g_i(k)\left(\hat{k}\cdot\hat{u}_i\right)^2\right],\nonumber\\
g_i(k)&=&(3-2 \Delta)\epsilon_i\,\ln(k/ H),
\end{eqnarray}
where $\hat{u}_i$ are the unit vectors of the three different axes. We conclude that the invariance under dilations imply the presence of anisotropy in the
power spectrum (as well as in the higher correlation functions).


To make contact with the more standard way of expressing the 
anisotropic contribution to the power spectrum,  we assume that  the  anisotropy direction is along a generic   unit vector $\hat n$, which we can arbitrarily set to be the $x_3$-direction, and that in the orthogonal directions the expansion rate is the same. 
The metric (\ref{metricc}) becomes 

\beq
\d s^2=\frac{1}{H^2\eta^2}\left[\d \eta^2-  (-H\eta)^{-2\epsilon_1}(\d x_1^2+\d x_2^2)- (-H\eta)^{-2\epsilon_3}\d x_3^2\right].
\eeq
The average expansion rate $H$ and deviation from isotropy $\epsilon_H$ become 

\beq
H=\frac{1}{3}\left(2H_1+H_3\right)\,\,\,\,\,{\rm and} \,\,\,\,\,
\epsilon_H=\frac{2}{3}\frac{H_3-H_1}{\overline{H}}.
\eeq
With these definitions we have

\beq
\epsilon_1=\epsilon_2=-\frac{1}{2}\epsilon_H \,\,\,\,\,{\rm and} \,\,\,\,\,
\epsilon_3=\epsilon_H,
\eeq
and from Eq. (\ref{iu}) we get

\begin{eqnarray}
\label{iuu}
\langle\sigma_{\vk}\sigma_{-\vk}\rangle'&=& \frac{H^2}{2k^{3-2\Delta}}
+ (3-2\Delta)\frac{H^2}{2k^{5-2\Delta}}\ln(k/ H)\left[-\frac{1}{2}\epsilon_H k_1^2-\frac{1}{2}\epsilon_H k_2^2+\epsilon_H k_3^2\right]\nonumber\\
&=& \frac{\overline{H}^2}{2k^{3-2\Delta}}\left[1-\frac{3-2\Delta}{2}\epsilon_H\ln(k/\overline H)\right]+
\frac{3(3-2\Delta)}{2} \frac{H^2}{2k^{3-2\Delta}}\epsilon_H \ln(k/ H) \hat{k}^2_3\nonumber\\
&=& \frac{H^2}{2k^{3-2\Delta+(3-2\Delta)\epsilon_H/2}}+
\frac{3(3-2\Delta)}{2} \frac{H^2}{2k^{3-2\Delta}}\epsilon_H \ln(k/ H) \hat{k}^2_3.
\end{eqnarray}
It is important to stress at this point that the momenta in these expressions are referring to the Fourier transform with respect
to coordinates where rotational
invariance holds. This means that at the end of inflation at time $\eta_*$, all the momenta have to be properly rescaled as the coordinates in the metric do not exhibit rotational invariance. In other words, we have to rescale the coordinates in the following way

\beq
x_{1,2}\to  \frac{1}{(-H\eta_*)^{\epsilon_H/2}}\,x_{1,2}\,\,\,\,{\rm and}\,\,\,\, x_3\to (-H\eta_*)^{\epsilon_H}\,x_{3}.
\eeq
Performing this necessary rescaling 
and restoring  the arbitrary anisotropy direction along  $\hat n$, we finally get, denoting  by  $q=(-H\eta_*)k$  the physical
wavelength of the mode of interest at the end of inflation)

\begin{eqnarray}
\label{iuuu}
\langle\sigma_{\vec q }\sigma_{-\vec q}\rangle'&\simeq& \frac{H^2}{2k^{3-2\Delta+(3/2-\Delta)\epsilon_H}}\left[1+g(q)(\hat{k}\cdot\hat n)^2\right]\nonumber\\
g(q)&=&\frac{3(3-2\Delta)}{2}\epsilon_H\,\ln(q/ H).
\end{eqnarray}
Let us notice that we can cast the expression (\ref{iuuu}) in an alternative way by expanding the power spectrum to first-order in $ \epsilon_H$ (we rename the physical momentum now as $k$)

\begin{eqnarray}
\label{iuuu}
\langle\sigma_{\vec k }\sigma_{-\vec k}\rangle'&\simeq& \frac{H^2}{2k^{3-2\Delta}}\left[1+g_2(k)P_2(\hat{k}\cdot\hat n)\right]\nonumber\\
g_2(k)&=&(3-2\Delta)\epsilon_H\,\ln(k/ H),
\end{eqnarray}
where $P_2$ is the $\ell=2$ Legendre polynomials. In Appendix C we check this result with a direct computation.

 In this simple set-up the anisotropy is solely due to the anisotropic expansion 
of the universe. This was the case considered first in 
 the so-called ACW model \cite{wise} where  the anisotropic de Sitter expansion was due to a vector field $B_\mu$ which, by a Lagrange constraint,  satisfies the relation $B_\mu B^\mu=m^2$ and has a vacuum expectation value along the $x_3$-direction. On the top there is an isotropic  vacuum energy. The ACW model
suffers  of  instabilities related to the negative energy of the longitudinal mode of the vector field, but we have considered it just for the sake of comparison.

 We will start from expression (\ref{iuuu}) to elaborate our considerations about the correspondence between anisotropic four-dimensional de Sitter spacetimes and a three-dimensional boundary theory.
%

\subsection{Anisotropic de Sitter and its three-dimensional dual perspective}
When  the de Sitter spacetime is anisotropic, the isometry of the  metric is not any longer SO(1,4). Nevertheless, dilations are still isometries. As a consequence, the power spectrum and the higher correlator functions acquire an anisotropic dependence.
We would like to show in this subsection that this result has a dual interpretation: the four-dimensional anisotropic de Sitter spacetime
is in correspondence with an isotropic  three-dimensional boundary enjoying dilation symmetry. 
In this picture,  
the angular anisotropic dependence of the spectrum can be attributed to a non-zero 
expectation value of the stress tensor $\langle T_{ij} \rangle$ of the three-dimensional dual theory. Our procedure follows the one nicely introduced by Cardy   for anisotropic corrections to correlation functions in conformal systems \cite{cardy}.

Let us  consider a three-dimensional theory enjoying dilation symmetry and 
 recall that the stress tensor is determined by the response of the action under a general coordinate transformation. Indeed, under the transformation $x^i\to x^i+\alpha^i(\vx)$, the action $S$ changes as

\begin{eqnarray}
 \delta S=-\frac{1}{4\pi}\int \d^3x\,\partial^{(i} \alpha^{j)} \, T_{ij}.        \label{dS}
 \end{eqnarray} 
For  dilations we have $\alpha^i=\lambda x^i$, and hence invariance of the action simply means that the stress tensor is traceless $T^i_{\,\,\, i}=0$. In fact, the stress tensor may be viewed as the generator of  dilation transformations in the following sense. Consider a 
coordinate transformation $x^i\to x^i+\alpha^i(\vx)$. We assume that, with $\epsilon_{\rm in}<\epsilon_{\rm out}$, inside the ball 
$|\vec{x}|<\epsilon_{\rm in}$ this transformation is a dilation, {\it i.e.} $\alpha^i=\lambda x^i$, outside the ball 
$|\vec{x}|>\epsilon_{\rm out}$, $\alpha^i=0$, and in between $\epsilon_{\rm in}<|\vec{x}|<\epsilon_{\rm in}$, $\alpha^i$ is a general differentiable function. 
\begin{figure}[h!]
\vskip -4cm
    \begin{center}
      \includegraphics[scale=.4]{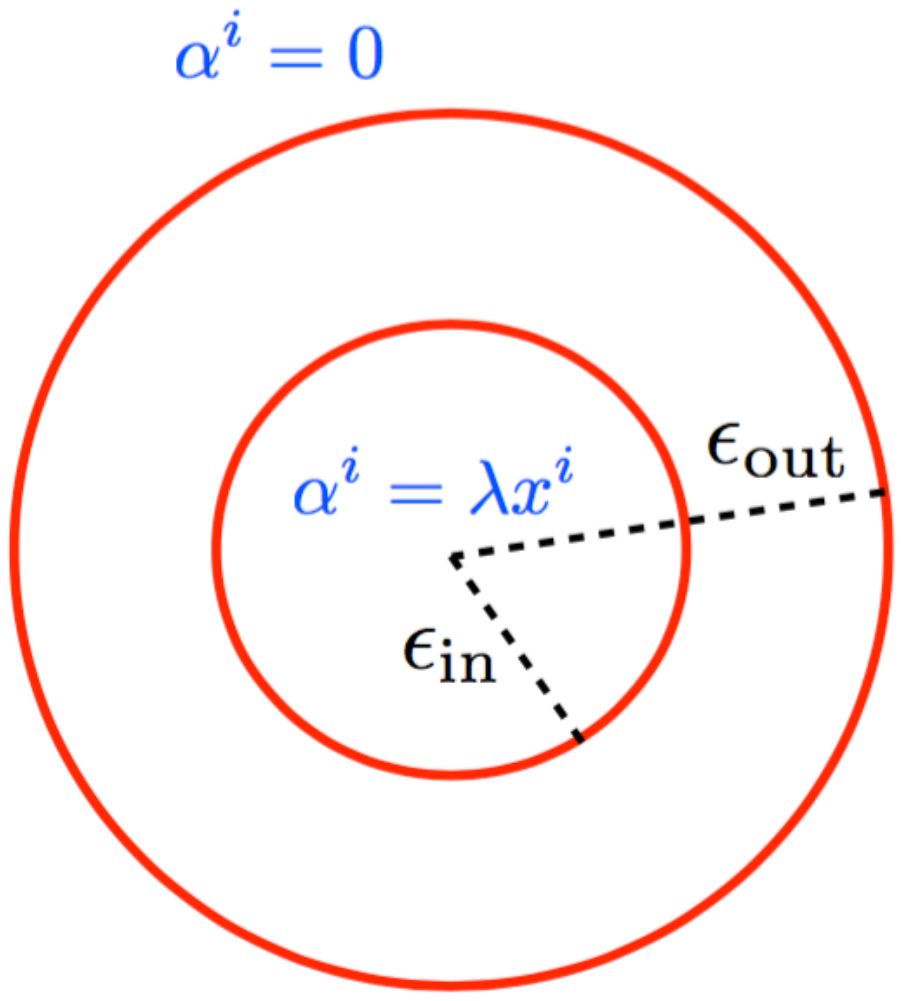}
    \end{center}
     \vskip -4cm
     \caption{\footnotesize A schematic view of the transformation of the coordinates adopted in the text.}  
\end{figure}
This transformation induces a corresponding transformation to the action $S\to S+\delta S$ where, using  Eq. (\ref{dS}) and the conservation of the stress tensor 
$\partial^iT_{ij}=0$,

\begin{eqnarray}
\delta S=-
\frac{1}{4\pi}\int\d^3 x\, \partial^i\left(\alpha^i T_{ij}\right)=\frac{1}{4\pi}
\int_{|\vec{x}|=\epsilon_{\rm out} }\d\Omega_2\,\alpha^i T_{ij}x^j|\vec{x}|-\frac{1}{4\pi}\int_{|\vec{x}|=\epsilon_{\rm in}}\d\Omega_2\, \alpha^i T_{ij}x^j|\vec{x}|,
\end{eqnarray}
where $\d\Omega_2$ is the integral measure on the unit two-dimensional sphere. 
The surface integral at $|\vec{x}|=\epsilon_{\rm out} $ vanishes due to the fact that $\alpha^i=0$ by continuity there, and so we are left with 

\begin{eqnarray}
 \delta S=-
\frac{1}{4\pi}\int_{|\vec{x}|=\epsilon_{\rm in}}\d\Omega_2\, \alpha^i T_{ij}x^j|\vec{x}|.
\end{eqnarray}
We are now interested in the response to the transformation 
$x^i\to x^i+\alpha^i(\vx)$ of the correlator of a field $\pi(0)$ of scaling dimension $\df$ with  operators ${\cal O}_1(\x_1) {\cal O}_2(\x_2)\cdots {\cal O}(\x_n)$
of arbitrary spin and dimension at $x_a^i$ ($a=1,\cdots, n$) with $|\vec{x}_a|>\epsilon_{\rm out}$. The  boundary field $\pi(\vx)$ of scaling dimension $h=3-\Delta$ is supposed to be the dual of the free scalar bulk scalar field $\sigma$ with dimension $\Delta$.

We find that 
\begin{eqnarray}
\delta\expect{\pi(\0){\cal O}_1(\x_1)\cdots{\cal O}_n(\x_n)}=\expect{\delta S\,\pi(\0){\cal O}_1(\x_1)\cdots{\cal O}_n(\x_n)}. 
\end{eqnarray}
Since, only $\pi(0)$ transforms whereas all other fields ${\cal O}_n(\x_n)$ do not change ($\alpha^i=0$ at $x_n^i$, and thus 
$\delta {\cal O}_n=0$), we get

\begin{eqnarray}
\delta\pi(0)=-\frac{1}{4\pi}\int_{|\vec{x}|=\epsilon_{\rm in}} \, \d\Omega_2\,\alpha^i T_{ij}\,x^j|\,\vec{x}|\pi(0). \label{dil}
\end{eqnarray}
This form of the transformation and the fact that the stress tensor is symmetric, traceless and has scaling dimension three are enough to determine part of the operator product expansion (OPE) of $T_{ij}(x)$ with $\pi(0)$. The latter  will contain among others the term

\begin{eqnarray}
 T_{ij}(\vx)\pi(0)=\frac{c_1}{|\vec{x}|^3}t_{ij}(x)\pi(0)+\cdots, \label{ope}
 \end{eqnarray} 
 where 
 
 \begin{eqnarray}
t_{ij}(\x)=\frac{1}{|\vec{x}|^2}\left(x_ix_j-
\frac{1}{3}|\vec{x}|^2\delta_{ij}\right). \label{tff}
\end{eqnarray}
Since, the dimension of $\pi$ is $\df$, we have for a dilation $\alpha^i=\lambda x^i$, 

\begin{eqnarray}
\delta \pi(0)=-\lambda \df\,  \pi(0).
\end{eqnarray}
On the other hand, using the OPE of Eq. (\ref{ope}) in Eq. (\ref{dil}), we find 

\begin{eqnarray}
\df=\frac{c_1}{4\pi}\int_{|\vec{x}|=\epsilon_{\rm in}}\d\Omega_2\, \frac{x^ix^j}{|\vec{x}|^4}
\left(x_i x_j-\frac{1}{3}|\vec{x}|^2\delta_{ij}\right),
\end{eqnarray}
from which it follows  

\begin{eqnarray}
c_1=\frac{3\df}{2}. 
\end{eqnarray}
In order to find the induced  angular dependence of the spectrum due to a non-zero vacuum expectation value of the stress tensor, we should consider the contribution of the latter to the OPE of the field $\pi$ with itself. Since, the dimensions of $\pi$ and $T_{ij}$  are $\df$ and  three, respectively, the OPE of  a scalar field 
$\pi(\vec{x})$ with itself will have the form

\begin{eqnarray}
\pi(\vec{x})\pi(0)\propto \frac{1}{|\vec{x}|^{2\Delta}}+C^{ij}(\vec{x})T_{ij}(0)+\cdots,  \label{ff}
\end{eqnarray}
where 

\begin{eqnarray}
C^{ij}(\x)= \frac{c_\sigma}{|\vec{x}|^{2\df-3}} t^{ij}(\x). 
\end{eqnarray}
The parameter $c_\sigma$ is not arbitrary,  but it is specified by $c_1$ and the central charge $c_{\rm T}$ defined in the two-function of the energy-momentum tensor

\begin{eqnarray}
\expect{T_{ij}(\x)T_{mn}(\0)}=\frac{c_{\rm T}}{|\vec{x}|^6}
\left({\cal I}_{im}(\x){\cal I}_{jn}(\x)+
{\cal I}_{in}(\x){\cal I}_{jm}(\x)-\frac{2}{3}\delta_{ij}\delta_{mn}\right), \label{tt}
\end{eqnarray}
where 
\begin{eqnarray}
{\cal I}_{ij}(\x)=\delta_{ij}-2\frac{x^ix^j}{|\vec{x}|^2}.
\end{eqnarray}
In addition, the three-function $\expect{T_{ij}\pi\pi}$ is given by 

\begin{eqnarray}
\expect{T_{ij}(\x_1)\pi(\x_2)\pi(\x_3)}=
\frac{c_1}{x_{12}^3x_{23}^{2\df-3}x_{31}^3}{\cal I}_{ik}(x_{13}){\cal I}_{jn}(x_{13})t_{kn}(x_{12}),
\end{eqnarray}
where $x_{ij}=|\x_i-\x_j|$. For $\x_3=0$, $\x_1=\x$ and $\x_2=
\vec{y}$ with $|\vec{y}|\to 0$, the leading behaviour of Eq. (\ref{tff}) turns out to be

\begin{eqnarray}
\expect{T_{ij}(\x)\pi(\vec{y})\pi(0)}=
\frac{c_1}{|\vx|^6|\vec{y}|^{2\df-3}}{\cal I}_{ik}(\x){\cal I}_{jn}(\x)t_{kn}(\vec{y}), \label{tf0}
\end{eqnarray}
which should be equal, after using Eq. (\ref{ff}), to 
\begin{eqnarray}
C^{mn}(\vec{y})\expect{T_{ij}(\x)T_{mn}(0)}. \label{ctt}
\end{eqnarray}
By comparing, Eqs. (\ref{ctt}),  (\ref{tf0}) and  (\ref{tt}) one finally finds   \cite{cardy}

\begin{eqnarray}
c_\sigma=\frac{3\df}{2c_{\rm T}}. 
\end{eqnarray}
The OPE is may be written in  momentum space as

\begin{eqnarray}
\pi_{\vec{k}}\pi_{-\vec{k}}\propto \frac{1}{k^{3-2\df}}+C(k)\left(k^ik^j-\frac{1}{3}k^2\delta^{ij}\right) T_{ij}(0)+\cdots,
\end{eqnarray}
and therefore
\begin{eqnarray}
\langle\pi_{\vec{k}}\pi_{-\vec{k}}\rangle'\propto
\frac{1}{k^{3-2\df}}+C(k)\left(k^ik^j-\frac{1}{3}k^2\delta^{ij}\right) \langle
T_{ij}(0)\rangle+\cdots. \label{2pt}
\end{eqnarray}
For an anisotropy along a direction specified by the unit vector $\vec{n}$, the only possibility for the expectation value of the energy-momentum tensor is 

\begin{eqnarray}
\langle
T_{ij}(0)\rangle=A \, \left(n_in_j-\frac{1}{3}\delta_{ij}\right), 
\end{eqnarray}
where $A$ is a constant. Since 

\begin{eqnarray}
\left(k^ik^j-\frac{1}{3}k^2\delta^{ij}\right)\left(n_in_j-\frac{1}{3}\delta_{ij}\right)=\frac{2}{3}k^2P_2(\hat{k}\cdot\hat{n}), 
\end{eqnarray}
we may write Eq. (\ref{2pt}) as 

\begin{eqnarray}
\langle\pi_{\vec{k}}\pi_{-\vec{k}}\rangle'\propto
\frac{1}{k^{3-2\df}}\left[1+\widetilde{C}(k)\,P_2(\hat{k}\cdot\hat{n})\right]+\cdots. \label{22pt}
\end{eqnarray}
with 
\begin{eqnarray}
\widetilde{C}(k)=\frac{16}{3}A(\df-1)(\df-2)(\df-3)\cot(\pi \df)k^2C(k).
\end{eqnarray}
Eq. (\ref{22pt}) reproduces correctly the angular dependence of the spectrum found in Eq. (\ref{iuuu}). In addition, the function $\tilde C(k)$ can be found by solving the Ward identity corresponding to the dilation symmetry

\begin{eqnarray}
\label{a}
\left[3-2\df +\left(1-\frac{\epsilon_H}{2}\right)(k_1\partial_{k_1}
+k_2\partial_{k_2}+(1+\epsilon_H)k_3\partial_{k_3}\right]\langle\pi_{\vk}\pi_{-\vk}\rangle'=0.
\end{eqnarray}
 As a consequence $\widetilde{C}(k)$ satisfies the equation

\begin{eqnarray}
\label{aa}
(3-2\df)\, \epsilon_H\, k^2\left(1-3(\hat{k}\cdot\hat{n})^2\right)+2k^3\, P_2(\hat{k}\cdot\hat{n})\frac{\d\widetilde{C}(k)}{\d k}=0. \label{soo}
\end{eqnarray}
which has the solution

\begin{eqnarray}
\widetilde{C}(k)= \frac{(3-2\df)}{k^{3-2\df}}\, \epsilon_H\, \ln (k/H).\label{ckk}
\end{eqnarray}
In the dual picture the field  $\pi$ is a dual operator in the putative three-dimensional dual theory of the field $\sigma$. In the wavefunction of the universe approach \cite{mal} and in the Gaussian approximation this means that such a wavefunction is

\be
\Psi[\sigma]\sim {\rm exp}\left[-\frac{1}{2}\int\frac{\d^3k}{(2\pi)^3}\langle\pi_{\vk}\pi_{-\vk}\rangle'\sigma^2\right].
\ee
This means that the two-point correlator of the bulk field $\sigma$ is given by

\begin{eqnarray}
\langle\sigma_{\vk}\sigma_{-\vk}\rangle'&=&\frac{k^{3-2h}}{2\left[1+(3-2\df)\epsilon_H\, \ln (k/H)\right]}\nonumber\\
&\simeq&\frac{1}{2k^{3-2\Delta}}\left[1+(3-2\Delta)\epsilon_H\, \ln (k/H)\right],
\end{eqnarray}
where we have used the relation $h=3-\Delta$ and expanded for small anisotropies. This result
exactly matches 
the result in  Eq. (\ref{iuuu}).

We can also examine the case in which the anisotropy is generated by the vacuum expectation value  of a generic spin $\ell$ operator ${\cal{O}}_{i_1i_2\cdots i_\ell}$. Indeed, let us recall that the OPE of a scalar field takes the form \cite{OPE}
\be
\pi(\vx)\pi(0)
\propto
\sum_{{\cal O}}C_\ell\frac{x^{i_1}x^{i_2}
\cdots x^{i_\ell}}{|\vec{x}|^{2\Delta-\Delta_{\cal O}+n}}\, \Phi\!\left(\!\frac{1}{2}(\Delta_{\cal O}+n);\Delta_{\cal O}\!+\!n;\x\cdot \vec{\partial}\right){\cal{O}}_{i_1i_2\cdots i_\ell}(0), \label{OPEG}
\ee
where $\Phi(a;b;z)$ is the confluent hypergeometric function and 
${\cal{O}}_{i_1i_2\cdots i_\ell}$ is a spin $\ell$ operator ({\it i.e.}, symmetric and traceless). By using the expansion 

\begin{eqnarray}
\Phi(a;b;z)=1+\frac{a}{b}z+\frac{a(1+a)}{2b(1+b)}z^2+\cdots, 
\end{eqnarray}
we find that the leading contribution of the spin $\ell$ operators to the two-function in momentum space will take the general form

\begin{eqnarray}
\langle\pi_{\vec{k}}\pi_{-\vec{k}}\rangle\propto
\frac{1}{k^{3-2\df}}+C_\ell(k)\frac{k^{i_1}k^{i_2}
\cdots k^{i_\ell}}{k^{3-2\Delta+\Delta_{\cal O}-n}}
\expect{{\cal{O}}_{i_1i_2\cdots i_\ell}(0)}+\cdots\, . \label{2ptg}
\end{eqnarray}
%
%
In an anisotropic background specified by a unit vector $n^i$, 
the expectation value of a spin $\ell$ operators  can take the form

\begin{eqnarray}
\widetilde{{\cal{O}}}_{i_1i_2\cdots i_\ell}=A_\ell\big{[}n_{i_1}\cdots n_{i_\ell}\big{]},
\end{eqnarray}
where $A_\ell$ is a constant and $\big{[}n_{i_1}\cdots n_{i_n}\big{]}$ denotes  the traceless symmetric part of the polynomial $n_{i_1}\cdots n_{i_n}$. For example,

\begin{eqnarray}
&&\big{[}n_{i_1}n_{i_2}\big{]}=n_{i_1}n_{i_2}-\frac{1}{3}\delta_{i_1i_2}, \nonumber \\
&&\big{[}n_{i_1}n_{i_2}n_{i_3}\big{]}=n_{i_1}n_{i_2}n_{i_3}-\frac{1}{5}\left(n_{i_1}\delta_{i_2i_3}+n_{i_2}\delta_{i_1i_3}+n_{i_3}\delta_{i_1i_2}\right),
\end{eqnarray}
and so on. Using the relation

\begin{eqnarray}
\hat{k}^{i_1}\hat{k}^{i_2}
\cdots \hat{k}^{i_\ell}\big{[}n_{i_1}\cdots n_{i_\ell}\big{]}=
\frac{2^\ell(\ell!)^2}{(2\ell)!}P_\ell(\hat{k}\cdot{n}),
\end{eqnarray}
where $P_\ell$  are the Legendre polynomials, we get that the two-function can be written as 

\begin{eqnarray}
\langle\pi_{\vec{k}}\pi_{-\vec{k}}\rangle'\propto
\frac{1}{k^{3-2\df}}\left(1+\widetilde{C}_\ell(k)P_\ell(\hat{k}\cdot{n})
+\cdots\right). \label{2ptg}
\end{eqnarray}
Using again the Ward identity  we find that $\widetilde{C}_\ell(k)$ satisfies the equation

\begin{eqnarray}
(3-2\df)\, \epsilon_H\, k^2\left(1-3(\hat{k}\cdot\hat{n})^2\right)+2k^3\, P_\ell(\hat{k}\cdot\hat{n})\frac{\d\widetilde{C}_\ell(k)}{\d k}=0. \label{soo}
\end{eqnarray}
A solution to this equation  exists only for $\ell=2$, which  corresponds to  a non-zero vacuum expectation value of the stress tensor as discussed above.  In other words, the only vacuum expectation value that it is consistent with scale invariance in the anisotropic de Sitter spacetime is the one  of a spin $\ell=2$ operator.

\section{Conclusions}
In this paper we have discussed several topics related to inflation and de Sitter spacetimes from a three-dimensional perspective
by making use of the properties of radially quantized CFT$_3$. Despite the fact that some of the results, {\it e.g.} on the Higuchi bound, are not totally
new, we have offered  new interpretations and tools which might turn out to be useful in addressing other topics.   

One interesting question we are investigating is if, under some assumptions,
 the convergence of the operator product expansion  and conformal block decomposition in unitary CFT$_3$ \cite{rattazzi} may deliver an upper bound
 on four-point correlators. This will be quite exciting  because it will have a strong impact on the next future experimental efforts
 of detecting  primordial four-point functions.
 
 Another issue worth understanding is the following.
 As already pointed out by in Ref.  \cite{strominger},  in the dS/CFT correspondence the  dual CFT$_3$ may be non-unitary.  This   happens when  there are sufficiently massive stable scalars which correspond to  complex conformal weights.
While in inflation we are mainly interested
in very light degrees of freedom and this issue might not represent a real problem, it represents  nevertheless a strong motivation to study the dS/CFT$_3$ in more detail. 

Unitarity, or reflection positivity,  of the CFT$_3$ is the property we made use of   to derive the Higuchi bound. 
Since the bound corresponds to real scaling dimensions the derivation is consistent. 
In AdS/CFT   there  Lorentzian theories on the boundary and in the bulk. There is time evolution and this implies that if one theory  is unitary, the same must be true for the other. In dS/CFT$_3$ the  equivalence is between an Euclidean theory (future boundary) and a Lorentzian theory (bulk).  No ``real" time evolution in the Euclidean theory is present and so there is no right to require    unitary a priori.  Nevertheless,  if $\mathscr{H}_{\rm cyl}$
has arbitrarily large and negative eigenvalues, then the Hamiltonian  is unbounded and one is obliged to  consider states that have no overlap with arbitrarily high-energy eigenstates.

 Finally, it will be interesting to investigate more in details the role played by symmetries in anisotropic inflation. We hope to come back to these issues soon.

\section*{Acknowledgments}
We thank M. Sloth and J. Sonner  for comments. A.R. is supported by the Swiss National Science Foundation (SNSF), project {\sl Investigating the
Nature of Dark Matter}, project number: 200020-159223.

\appendix
\numberwithin{equation}{section}
\section{An alternative derivation of the relation between mass and scaling dimension for scalar fields}
Another, maybe more intuitive way to obtain the relation (\ref{MH}) is to consider a massive scalar field with action

\beq
S=\int\d^3x\d\eta\sqrt{-g}\,\left(-\frac{1}{2}(\partial\phi)^2-\frac{1}{2}m^2\phi^2\right)=\int\d^3x\d\eta\, a^4\left[\frac{1}{2a^2}(\dot{\phi}^2-(\nabla\phi)^2)-\frac{1}{2}m^2\phi^2\right],\label{s}
\eeq
where $a(\eta)=-1/H\eta$. By integrating by parts one obtains

\beq
S=-\frac{1}{2}\int\d^3x\d\eta\, a^2\phi \left[\ddot\phi+2\frac{\dot a}{a}\dot\phi+m^2a^2\phi\right]+\frac{1}{2}\int\d^3x\d\eta\, a^2(\nabla\phi)^2.
\eeq
By posing $\phi(\eta,\vx)=\eta^\D\varphi(\vx)$, the action becomes

\beq
S=-\frac{1}{2}\int\d^3x\d\eta\, \frac{1}{H^2\eta^{4-2\D}}\varphi \left(\D(\D-3)+\frac{m^2}{H^2}\right)\varphi+\frac{1}{2}\int\d^3x\d\eta\, \frac{1}{H^2\eta^{2-2\D}}(\nabla\varphi)^2.
\eeq
If the coefficient of the first piece vanishes one is left with a free massless scalar field and therefore with a conformal field theory on ${\mathbb R}^3$. This corresponds precisely to the condition (\ref{MH}).

\section{Calculation of CFT correlation functions  through the state-operator correspondence}
We present here an alternative  proof that correlation functions on conformally flat backgrounds  can be calculated by rescaling the flat space correlation functions. It makes use of   the state-operator correspondence.  
One has 
\beq
\zeta(\vx_1)|0\rangle=e^{i\vec P\cdot\vx_1}|\D_\zeta\rangle,
\eeq
and

\beq
\langle 0|\zeta(\vx_2)=r_2^{-2\D_\zeta}\langle 0|\left[\zeta({\cal I}\vx_2)\right]^\dagger=r_2^{-2\D_\zeta}\langle \D_\zeta| e^{-i\vec K\cdot{\cal I}\vx_2}.
\eeq
The two-point correlator of a scalar field with scaling dimension $\D_\zeta$  becomes

\beq
\langle 0|\zeta(\vx_2)\zeta(\vx_1)|0\rangle=r_2^{-2\D_\zeta}\langle\D_\zeta| e^{-i\vec K\cdot {\cal I}\vx_2}e^{i\vec P\cdot\vx_1}|\D_\zeta\rangle=
\frac{1}{r_1^{\D_\zeta}}\frac{1}{r_2^{\D_\zeta}}\sum_m\langle m,\vec{n}_2|m,\vec{n}_1\rangle\left(\frac{r_1}{r_2}\right)^{\D_\zeta+m},
\eeq
where
\beq
|m,\vec{n}\rangle=\frac{1}{m!}(\vec P\cdot \vec n)^m |\D_\zeta\rangle\,\,\,\,{\rm and}\,\,\,\,\langle m,\vec{n}|=\frac{1}{m!}\langle\D_\zeta|(\vec K\cdot \vec n)^m,
\eeq
and we have taken into account that all the  cross-terms with different powers of $\vec K$ and $\vec P$ give 
vanishing matrix elements as states with different energies are orthogonal.
For  $m=1$ for instance, we 
have 

\begin{eqnarray}
\langle\D_\zeta| K_i P_j|\D_\zeta\rangle&=&\langle\D_\zeta| [K_i,P_j]|\D_\zeta\rangle+\langle\D_\zeta| P_j K_i|\D_\zeta\rangle=\langle\D_\zeta|2i(D\delta_{ij}-L_{ij})|\D_\zeta\rangle
\nonumber\\
&=&2\D_\zeta\delta_{ij}\langle\D_\zeta|\D_\zeta\rangle=2\D_\zeta\delta_{ij},
\end{eqnarray}
and similarly for the higher terms. The important point though is that a rescaling $r\to e^\lambda r$ does not alter neither the scalar products nor
the ratio $(r_1/r_2)^{\D_\zeta+m}$. It appears only on the overall terms $(r_1 r_2)^{-\D_\zeta}$.
 \section{Anisotropy in the power spectrum: direct computation}
We reobtain the generic  result (\ref{iuuu}) with a more explicit computation. Consider a free   massless scalar field $\sigma$ (therefore from now  $\Delta=0$) in the background metric 

\begin{eqnarray}
\d s^2=\frac{1}{H^2\eta^2}\left[-\d\eta^2+(-H\eta)^{\epsilon_H}(\d x_1^2+\d x_2^2)+
\left(-H\eta\right)^{-2\epsilon_H}\d x_3^2\right]
\end{eqnarray}
After redefining the scalar field $\sigma$ as 

\begin{eqnarray}
\sigma=\frac{\chi}{H\eta},
\end{eqnarray}
we find that  the equation of motion for the Fourier modes $\chi_{\vk}(\eta)$ turns out to be (primes indicate derivatives with respect to the conformal time)

\begin{eqnarray}
\chi_{\vk}''+\left[-\frac{2}{\eta^2}+(-{H} \eta)^{-\epsilon_H} 
k_\perp^2+(-{H} \eta)^{2\epsilon_H}k_3^2\right]\chi_{\vk}=0,
\end{eqnarray}
where $k_\perp^2=k_1^2+k_2^2$. 
Expanding in $\epsilon_H$, we find that to first order we have

\begin{eqnarray}
\chi_{\vk}''+\left(k^2-\frac{2}{\eta^2}\right)\chi_{\vk}-\epsilon_H\ln(-H \eta) \left(k^2-3k_3^2\right)\chi_{\vk}=0,
\end{eqnarray}
where $k^2=k_1^2+k_2^2+k_3^2$.
Looking for solutions of the form
\begin{eqnarray}
\chi_{\vk}(\tau)=\chi_{\vk}^{(0)}(\eta+\epsilon_H \chi_{\vk}^{(1)}(\eta),
\end{eqnarray}
we find  that $\chi_{\vk}^{(0)}(\eta)$ satisfies the standard equation in de Sitter

\begin{eqnarray}
{\chi_{\vk}^{(0)}}''+\left(k^2-\frac{2}{\eta^2}\right)\chi_{\vk}^{(0)}=0,
\end{eqnarray}
 and the appropriately normalized solution is 

\begin{eqnarray}
\chi_{\vk}^{(0)}(\eta)=\frac{1}{\sqrt{2k}}e^{-ik \eta}
\left(1-\frac{i}{k\eta}\right).
\end{eqnarray}
The field $\chi_k^{(1)}(\tau)$ satisfies 

\begin{eqnarray}
{\chi_{\vk}^{(1)}}''+\left(k^2-{H}^2\right)\chi_{\vk}^{(1)}=\ln(-{H}\eta) \left(k^2-3k_3^2\right)\chi_{\vk}^{(0)},
\end{eqnarray}
whose solution is

\begin{eqnarray}
{\chi_k^{(1)}}&=&\frac{k^2-3k_3^2}{4 \sqrt{2}
   k^{7/2} \eta }\bigg\{ e^{-i k \tau } \Big{(}3 e^{2 i k \eta } (k
   \tau +i) \text{Ei}(-2 i k \eta )+\left(2 i k^2 \eta ^2
   +3 k \eta -3 i\right)
   \ln (-{H} \eta )
   \nonumber \\
   &&\qquad \qquad-2 i \left(k^2 \eta ^2-i k \eta +2\right)\Big{)}\bigg\},
\end{eqnarray}
where $E_i(x)=-\int_{-x}^{\infty} e^{-t}\d t/t$ is exponential integral.
In the limit $-k\eta\ll 1$, the solution is given in the 
leading $\ln(k/H)$ order 

\begin{eqnarray}
\chi_k^{(1)}\approx\frac{3i}{4\sqrt{2}}\frac{(k^2-3k_3^2)}{k^{7/2}\eta}\ln (k/H).
\end{eqnarray}
The final power spectrum is therefore

\begin{eqnarray}
\langle\sigma_{\vec k }\sigma_{-\vec k}\rangle'&=&\frac{{H}^2}{2k^3}\left\{1+\frac{3}{4}\epsilon_H \big{[}3(\hat{k}\cdot n)^2-1\big{]}\ln(k/\overline{H}\right\}\nonumber 
\\
&=&\frac{{H}^2}{2k^3}
\left(1+3\epsilon_H\ln(k/\overline{H})\, P_2(\hat{k}\cdot n)\right),
\end{eqnarray}
which coincides with the result (\ref{iuuu}) obtained by symmetry arguments.

\end{document}